\documentclass[12pt]{article}
\usepackage[T1]{fontenc}
\usepackage[bbgreekl]{mathbbol}
\usepackage{mathtools}
\usepackage{amssymb,amsmath,mathbbol,mathrsfs}
\DeclareSymbolFontAlphabet{\mathbbm}{bbold}
\DeclareSymbolFontAlphabet{\mathbb}{AMSb}%
\usepackage[mathscr]{euscript}
\usepackage{bm}
\usepackage{xcolor}
\usepackage{tcolorbox}
\usepackage{enumitem}
\usepackage{dsfont}
\usepackage{slashed}
\usepackage{fullpage}
\usepackage{cite}
\usepackage{currvita}
\usepackage[colorlinks=true,linkcolor=blue,citecolor=magenta,linktocpage=true]{hyperref}
\usepackage{braket}
\usepackage{titlesec}
\titleformat*{\section}{\normalsize\bfseries}
\titleformat*{\subsection}{\normalsize\bfseries}
\titleformat*{\subsubsection}{\normalsize\bfseries}
\makeatletter
\renewcommand{\@dotsep}{1000}
\newcommand{\be}{\begin{equation}}
\newcommand{\ee}{\end{equation}}
\newcommand{\bea}{\begin{eqnarray}}
\newcommand{\eea}{\end{eqnarray}}

\newcommand{\lb}{\label}
\renewcommand{\bar}{\overline}

\newcommand{\cL}{{\mathcal{L}}}

\renewcommand{\O}{\mathcal{O}}

\newcommand{\sE}{\mathscr{E}}
\newcommand{\sJ}{\mathscr{J}}
\newcommand{\sP}{\mathscr{P}}

\newcommand{\sS}{\mathscr{S}}
\newcommand{\sK}{\mathscr{K}}
\newcommand{\sD}{\mathscr{D}}

\renewcommand{\a}{{\alpha}}
\renewcommand{\b}{{\beta}}
\newcommand{\s}{\sigma}
\newcommand{\lam}{{\lambda}}

\newcommand{\mr}{\mathring}
\newcommand{\rd}{\mathrm{d}}
\newcommand{\e}{\mathrm{e}}

\newcommand{\sss}{\scriptscriptstyle}
\newcommand{\mrq}{\mathring{q}}
\newcommand{\mrD}{\mathring{\mathscr{D}}}

\newcommand{\pa}{\partial}

\newcommand{\two}{{}^{\sss (2)}}
\newcommand{\Carr}{{\sss \text{Carr}}}

\begin{document}

\title{\Large{\textbf{\sffamily Carrollian hydrodynamics from symmetries}}}
\author{\sffamily Laurent Freidel$^1$ \& Puttarak Jai-akson$^{1,2}$}
\date{\small{\textit{
$^1$Perimeter Institute for Theoretical Physics,\\ 31 Caroline Street North, Waterloo, Ontario, Canada N2L 2Y5\\
$^2$Department of Physics and Astronomy, University of Waterloo,\\ 200 University Avenue West, Waterloo, ON, N2L 3G1, Canada\\}}}

\maketitle

\begin{abstract}

In this work, we revisit Carrollian hydrodynamics, a type of non-Lorentzian hydrodynamics which has recently gained increasing attentions due to its underlying connection with dynamics of spacetime near null boundaries, and we aim at exploring symmetries associated with conservation laws of Carrollian fluids. With an elaborate construction of Carroll geometries, we generalize the Randers-Papapetrou metric by incorporating the fluid velocity field and the sub-leading components of the metric into our considerations and we argue that these two additional fields are compulsory phase space variables in the derivation of Carrollian hydrodynamics from symmetries. We then present a new notion of symmetry, called the near-Carrollian diffeomorphism, and demonstrate that this symmetry consistently yields a complete set of Carrollian hydrodynamic equations. Furthermore, due to the presence of the new phase space fields, our results thus generalize those already presented in the previous literatures. Lastly, the Noether charges associated with the near-Carrollian diffeomorphism and their time evolutions are also discussed.

\end{abstract}

\thispagestyle{empty}
\newpage
\setcounter{page}{1}

\hrule
\tableofcontents
\vspace{0.7cm}
\hrule


\section{Introduction}

The fascinating discovery journey of Carrollian physics has begun purely out of the mathematical curiosity of Lévy-Leblond \cite{Leblond1965} when he first proposed a new non-Lorentzian limit of  flat spacetime and derived its resulting contracted isometry group. The novel Carrollian\footnote{It was named after Lewis Carroll, the author of Through the Looking-Glass.} limit (also referred to as the ultra-relativistic limit and the ultra-local limit by different authors), lying at the opposite side to the familiar Galilean (non-relativistic) limit, along with associated geometries, symmetries, and rich physics unfolded in this limit have recently gained unprecedented attentions from many fields of theoretical physics, especially from the flat space holography community. 

Given any relativistic theory, non-Lorentzian variants are regarded 
as limits of the original relativistic theory as the \emph{speed of light}, $c$, approaches extreme values. There are two types of non-Lorentzian limit --- the Galilean limit and the Carrollian limit. The former corresponds to the limit $c \to \infty$\footnote{To be more rigorous, one would rather need to consider the dimensionless parameter $\frac{c}{v}$ where a characteristic velocity $v$ of a problem under consideration. The final results, however do not differ from naively using $c$ as the varying parameter.} while the latter corresponds to the opposite limit, $c \to 0$. Changing the speed of light affects spacetime structures, with a notable example being a structure of light cones. In the well-familiar Galilean case, light cones expand as $c \to \infty$ so that a free particle traverses spacetime without a speed limit, and there exists a notion of absolute time. Light cones, however collapse in the Carrollian limit $c \to 0$, hence freezing a free particle's motion and thereby completely inhibiting causal interactions between any spacetime events. It is in this sense that the Carrollian limit is sometimes called the ultra-local limit\footnote{Clarification of terminology is in order here. In terms of a dimensionless parameter $\frac{c}{v}$, the ultra-local limit corresponds to the case where $\frac{c}{v} \to 0$, meaning that the characteristic velocity of the problem tends to zero slower that $c$, in turn freezing the dynamics. On the other hand, the ultra-relativistic limit corresponds to the limit $\frac{c}{v} \to 1$, inferring that $v$ trends to $c$ in this limit. Unfortunately, these two terminologies have been mixed up and used interchangeably in the literature.}. In addition, the trademark of Carrollian theories, contrary to the Galilean case, is the existence of absolute space.
Spacetime symmetries are also contracted to the Galilei group and the Carroll group in their respective limits and their associated Lie algebras are derived from the Inönü-Wigner contraction \cite{Inonu1953}. 

Although Lévy-Leblond deemed practical utilization of the Carrollian limit and the Carroll group problematic, interest in Carrollian physics has recently been rejuvenated and gained ever-increasing attention due to its wealth of interesting aspects and applications. Developments in this topic include the generalization of Carroll geometries beyond flat spacetime \cite{Duval:2014uoa,Ciambelli:2019lap}. Non-trivial dynamics of systems of Carroll particles which occurs when turning on interactions and when particles are coupled to non-trivial background fields has been explored in \cite{Bergshoeff:2014jla, Marsot:2021tvq, Bidussi:2021nmp, Marsot:2022qkx}. Carrollian limit has also been studied in a wide range of relativistic theories: \cite{Gibbons:2002tv, Bagchi:2015nca, Bagchi:2016yyf, Bagchi:2017cte, Bagchi:2018wsn, Bagchi:2013bga, Bagchi:2021ban,Bergshoeff:2020xhv,Roychowdhury:2019aoi} for strings and branes, \cite{Ravera:2022buz} for supergravity theories, \cite{Basu:2018dub,Bagchi:2019clu, Banerjee:2020qjj} for electrodynamics, and aspects of the Carrollian gravity have also been addressed in \cite{Hartong:2015xda, Bergshoeff:2017btm, Duval:2017els, Morand:2018tke,Bergshoeff:2019ctr,Gomis:2019nih, Ballesteros:2019mxi, Gomis:2020wxp, Grumiller:2020elf, Hansen:2021fxi,Concha:2021jnn,Guerrieri:2021cdz,Perez:2021abf,Sengupta:2022jlx,Perez:2022jpr, Campoleoni:2022ebj}. Furthermore, a recent resurgence of Carrollian physics was largely catalyzed by the deep connection between Carroll geometries and null boundaries. At asymptotic null infinities, the connection between the (conformal) Carroll group and the Bondi-van der Burg-Metzner-Sachs (BMS) group \cite{Duval:2014uva,Duval:2014lpa} plays a central role in understanding of holography of asymptotically flat spacetime \cite{Bagchi:2022emh,Campoleoni:2022wmf,Bagchi:2022owq, Donnay:2022aba, Donnay:2022sdg} and thereby motivates the studies on Carrollian field theories \cite{Bagchi:2019xfx,Gupta:2020dtl,Bagchi:2022eav}. Carrollian physics has also appeared in the context of inflationary cosmology \cite{deBoer:2021jej}. 

In this work, we are interested in Carrollian fluids, which is another non-Lorentzian counterpart of relativistic fluids along with non-relativistic Galilean (or Navier-Stokes) fluids. Hydrodynamic equations governing the dynamics of Carrollian fluids are derived from the $c \to 0$ limit of the relativistic conservation laws of general relativistic fluids \cite{Ciambelli:2018xat}. While seemingly irrelevant to real-world fluids, Carrollian hydrodynamics has been shown to have applications in the field of black holes and holography \cite{Penna:2018gfx, Ciambelli:2018ojf, Ciambelli:2018wre, Campoleoni:2018ltl,Donnay:2019jiz,Ciambelli:2020eba,Ciambelli:2020ftk, Bagchi:2021qfe,Bagchi:2021gai,Petkou:2022bmz}. Akin to the Galilean case, the hydrodynamics equations of Carrollian fluids include the evolution equation of Carrollian energy density and the evolution equations of Carrollian momentum density (which are the Carrollian analog of the Navier-Stokes equations). One apparent difference between the two non-Lorentzian fluids lies in their respective continuity equations. In the Galilean case, there is a notion of a spin-0 quantity, the fluid mass density, which is conserved. Carrollian fluids instead exhibit a conserved spin-1 quantity, the Carrollian heat current. Carrollian hydrodynamics also has one more constraint equation.  

Since there are conservation laws for Carrollian fluids, a natural question followed by the Noether theorem then arises --- \emph{what symmetries are associated with these Carrollian conservation laws?} This question has already been addressed in \cite{Ciambelli:2018ojf, Petkou:2022bmz} where it has been demonstrated that Carrollian symmetry corresponds to the energy density and momentum density evolutions of Carrollian hydrodynamics. Thus, these works only managed to derive a part of the Carrollian fluid dynamics and the continuity equation of the Carrollian heat current and the constraint equation needed to be additionally supplemented. Our objective is to complete and hence generalize their results and provide a complete derivation of Carrollian hydrodynamics from symmetries. The incompleteness in their derivations that we are trying to fix stems from the following:
\begin{enumerate}[label = \roman*)]
\item Phase space of Carrollian hydrodynamics presented in \cite{Ciambelli:2018ojf, Petkou:2022bmz} lacked two fluid momenta, namely the Carrollian energy density and the sub-leading Carrollian viscous stress tensor, which appears in the constraint equation. We will show that these two momenta are conjugate to the fluid velocity field and the sub-leading sphere metric.

\item Carrollian symmetry is too restrictive. Because there are more equations of Carrollian hydrodynamics than those in the original relativistic hydrodynamics, more symmetries are required. The main result in this work is the enhanced symmetries, called the near-Carrollian symmetries, that yield all equations of Carrollian fluids.  
\end{enumerate}

The article is structured as follows. We start in section \ref{carroll-structure} with the introduction of Carroll structures, which serves as the most basic building block of Carroll geometries and Carrollian physics. We will discuss Carrollian hydrodynamics in section \ref{hydro} starting from relativistic conservation laws and then carefully consider the Carrollian limit. This closely follows the idea first explored in \cite{Ciambelli:2018xat} and we formalize it using the language of Carroll structures. Finally, in section \ref{hydro-sym}, we present a new viewpoint on Carrollian hydrodynamics based on symmetries. We propose a new notion of symmetries, which we call near-Carrollian symmetries, that extends the usual Carroll symmetries. We then demonstrate that these symmetries are associated to the full set of Carrollian hydrodynamics and derive the corresponding Noether charges. Lastly, we conclude in section \ref{conclusion} and comment about the possible avenue of investigations.

\section{Carroll Structures} \lb{carroll-structure}

We dedicate this section to describe \emph{universal} building blocks of Carroll geometries which underpin the research field of Carrollian physics: the so-called \emph{Carroll structures}. In what follows, we consider a $3$-dimensional space $H$ endowed with a null metric $q$ whose kernel is generated by a nowhere vanishing vector field $\ell$, meaning that $q(\ell, \cdot) =0$. The triplet $(H,\ell, q)$ provides a (weak) definition of Carroll structures\footnote{A strong definition of Carroll structure requires, in addition, an affine connection that parallel transports both the metric $q$ and the vector $\ell$ \cite{Duval:2014uoa,Duval:2014uva,Duval:2014lpa}. This connection however is not uniquely determined from the pair $(\ell, q)$ due to the non-degenerate nature of $q$.} \cite{Duval:2014uoa,Duval:2014uva,Duval:2014lpa, Ciambelli:2019lap}. Carroll structures are universal intrinsic structures of null surfaces both at finite distances \cite{Chandrasekaran:2018aop, Chandrasekaran:2021hxc, Ashtekar:2021wld}\footnote{\label{kappa} In \cite{Chandrasekaran:2018aop}, a complete universal structure of a null surface, viewed as a hypersurface embedded in an ambient spacetime, also includes an in-affinity function $\kappa$ of the null vector $\ell$. $\kappa$ is defined as a time connection which transforms under rescaling $\ell \to e^\alpha \ell$ as $ \kappa \to e^{\alpha}( \kappa + \ell[\alpha])$.} and at asymptotic infinities \cite{Ashtekar:2014zsa, Ashtekar:2018lor}.

Carroll structures are naturally described in the language of fiber bundle \cite{Ciambelli:2019lap}. This specifically means that the space $H$ is a fiber bundle, $p: H \to S$, with a 1-dimensional fiber. The $2$-dimensional base space $S$ can be chosen, for relevant physics at hand, to have a topology of a $2$-sphere (and will be dubbed the sphere in this article). We denote local coordinates on the sphere $S$ by $\{ \s^A \}$ and denote by $q_{AB} \bm{\rd} \s^A \circ \bm{\rd} \s^B$ a metric on $S$. 

Stemming from the fiber bundle structure of the space $H$, one defines the \emph{vertical subspace} of the tangent space $TH$, denoted by $\text{\bf vert}(H)$, to be a 1-dimensional kernel of the differential of the projection map, $\bm{\rd} p: TH \to TS$,
\begin{align}
\mathrm{\bf vert}(H) := \text{ker}(\bm{\rd} p).
\end{align}
 A vertical vector field $\ell \in \mathrm{\bf vert}(H)$ that belongs to the Carroll structure is a preferred representative of the equivalence class $[\ell]_{\sim}$ with the equivalence relation being the rescaling that preserve the direction of $\ell$, that is $\ell \sim \e^{\epsilon}\ell$, where $\epsilon$ is any arbitrary function on the space $H$. In this sense, the Carrollian vector $\ell$ also serves as a basis of the vertical subspace. Another element of the Carroll structure is a null Carrollian metric $q$ whose 1-dimensional kernel coincides with the vertical subspace, inferring that $q(\ell, \cdot) = 0$. The null metric can be obtained by pulling back a metric on the sphere $S$ by the projection map, that is
\begin{align}
q = p^* (q_{AB} \bm{\rd} \s^A \otimes \bm{\rd} \s^B) = q_{AB} \bm{e}^A \otimes \bm{e}^B,
\end{align}  
where we introduced the co-frame field $\bm{e}^A$ which is the pullback of the coordinate form $\bm{\rd} \s^A$ on the sphere $S$ by the projection map,  
\begin{align}
\bm{e}^A:= p^*(\bm{\rd} \sigma^A), \qquad \text{such that} \qquad \iota_\ell \bm{e}^A=0.
\end{align}
Note that the co-frame field, by definition, is a closed form on $H$, $\bm{\rd e}^A =0$. 

\begin{figure}[t]
\centering
\includegraphics[scale=0.25]{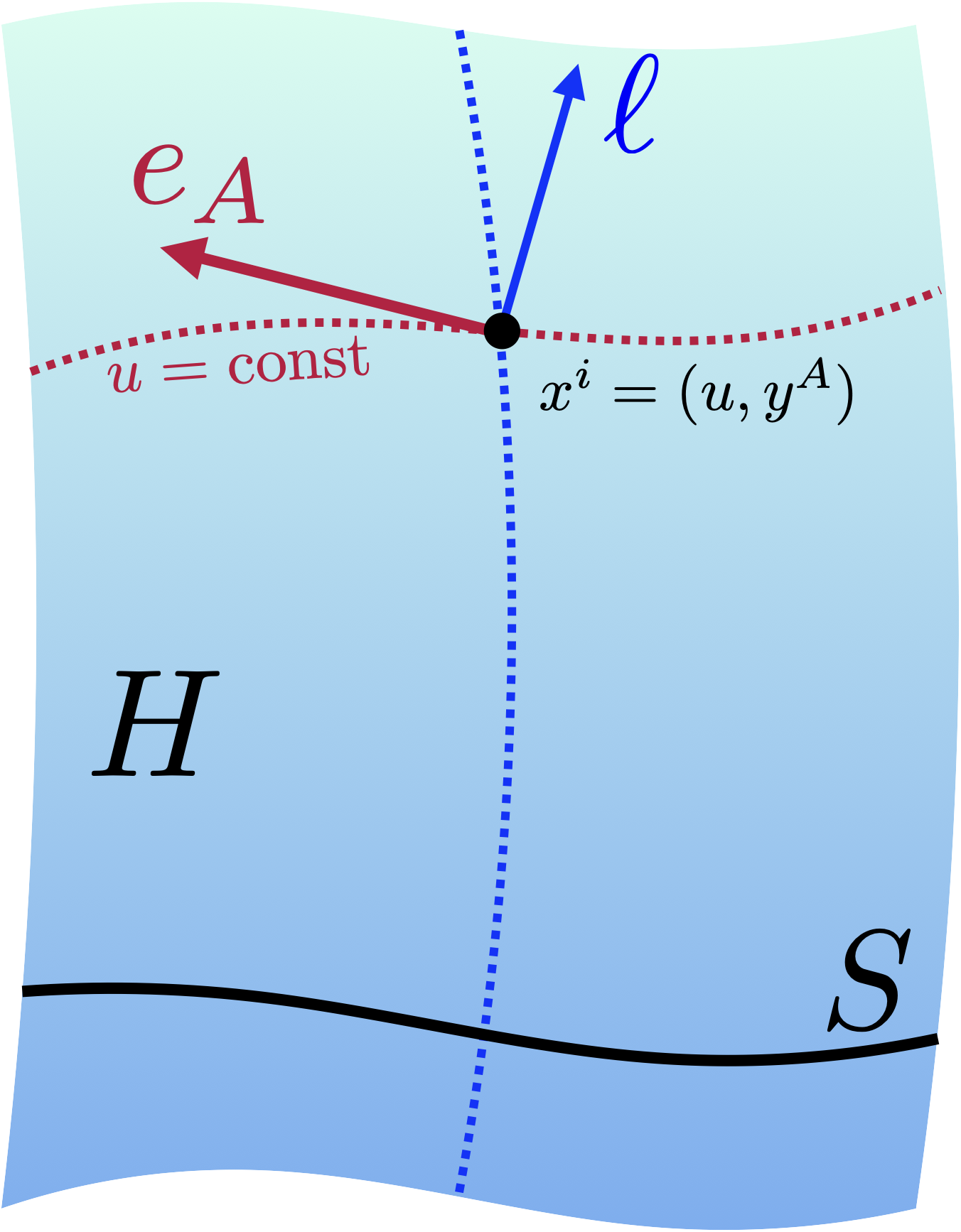} 
\caption{The space $H$ endowed with the Carroll structure. The general coordinates are $x^i = (u,y^A)$ where the surfaces at the cuts $u = \text{constant}$ are identified with the sphere $S$. The vertical vector $\ell$ and the horizontal vector $e_A$ span the tangent space $TH$} \lb{carroll-pic}
\end{figure}

Provided the Carroll structure on $H$, it then becomes possible to have an intrinsic separation of the tangent space $TH = \text{\bf vert}(H) \oplus \text{\bf hor}(H)$ into the aforementioned vertical subspace, $\text{\bf vert}(H)$, and its complement, the \emph{horizontal subspace} denoted by $\text{\bf hor}(H)$. This splitting can be achieved by introducing a connection 1-form, $\bm{k} \in T^*H$, dual to the vertical vector $\ell$,
\begin{align} 
\iota_{\ell} \bm{k} =1. \label{orthogonal}
\end{align}
The 1-form $\bm{k}$ is known as the \emph{Ehresmann connection} in the literature \cite{Ciambelli:2019lap,Chandrasekaran:2021hxc,Petkou:2022bmz }. Its kernel, seen as a linear map $\bm{k}: TH \to \mathbb{R}$, thus defines the $2$-dimensional horizontal subspace. This equivalently means that 
\begin{equation}
\text{\bf hor}(H) := \{ X \in TH | \iota_X \bm{k} =0 \}. 
\end{equation}  
In the following, we will denote a basis of the horizontal subspace by $e_A \in \mathrm{\bf hor}(H)$ which, by definition, obeys the condition $\iota_{e_A} \bm{k}=0$. Furthermore, without loss of generality, we can choose these horizontal basis vector fields to be ones that are dual to the co-frame field,
\begin{align}
\iota_{e_A} \bm{e}^B= \delta_A^B. 
\end{align}
The frame fields $(\ell, e_A)$ and the dual co-frame fields $(\bm{k}, \bm{e}^A)$ therefore serve as a complete basis for the tangent space $TH$ and the cotangent space $T^*H$, respectively (see Figure \ref{carroll-pic}). In this basis, any vector field $X \in TH$ and any 1-forms $\bm{\omega}\in T^* H$ can therefore be uniquely decomposed as follows:
\begin{align}
X= (\iota_X \bm{k}) \ell + (\iota_X \bm{e}^A) e_A, \qquad \text{and} \qquad \bm{\omega} = (\iota_{\ell} \bm{\omega}) \bm{k} + (\iota_{e_A} \bm{\omega}) \bm{e}^A.
\end{align}
Similarly, a differential of a function $F$ on the space $H$ can be expressed as  
\begin{align}
\bm{\rd} F = \ell [F]  \bm{k} + e_A[F] \bm{e}^A.
\end{align}

Lastly, having the intrinsic splitting of the tangent space $TH = \text{\bf vert}(H) \oplus \text{\bf hor}(H)$, one can naturally define the horizontal projector from the tangent space $TH$ to its horizontal components as
\begin{align}
q_i{}^j := e^A{}_ie_A{}^j = \delta^j_i -k_i \ell^j, \lb{C-projector}
\end{align}
and it satisfies the conditions $q_i{}^j k_j =0$ and $\ell^i q_i{}^j =0$.

\subsection{Acceleration, Vorticity, and Expansion}

Next, we introduce two important objects that are naturally inherited from the Carroll structure and they will later appear when discussing Carrollian hydrodynamics \cite{Ciambelli:2018xat, Ciambelli:2018ojf, Petkou:2022bmz}. These objects are the \textit{Carrollian acceleration}, denoted by $\varphi_A$, and the \textit{Carrollian vorticity}, denoted by $w_{AB}$. They are components of the curvature of the Ehresmann connection 1-form,
\begin{equation}
\begin{aligned}
\bm{\rd k} &:= -\left(\varphi_A \bm{k} \wedge  \bm{e}^A   +  \frac{1}{2} w_{AB} \bm{e}^A \wedge \bm{e}^B\right). \lb{d-k}
\end{aligned}
\end{equation}
Let us also recall that the co-frame $\bm{e}^A$ is closed, i.e., $\bm{\rd e}^A =0$. One can then show that the components $(\varphi_A,w_{AB})$ are also determined by the commutators of the basis vector fields. This correspondence can be established by invoking the identity $[\iota_X, \cL_Y ]\bm{\omega} = \iota_{[X,Y]} \bm{\omega}$ for any vector fields $X, Y \in TH$ and any 1-form $\bm{\omega} \in T^*H$. By making use of the Cartan formula, $\cL_X = \bm{\rd} \iota_X + \iota_X \bm{\rd}$, one can show that  
\begin{equation}
\iota_X \iota_Y \bm{\rd \o} = \iota_{[X,Y]}\bm{\o} + \cL_Y (\iota_X \bm{\o}) -\cL_X (\iota_Y \bm{\o}). \lb{duality-com}
\end{equation} 
Using this result and the property $\bm{\rd e}^A =0$, we show that the commutators of the frame fields satisfy the conditions,
\begin{align}
\iota_{[\ell, e_A]} \bm{e}^B = 0, \qquad \text{and} \qquad \iota_{[e_A, e_B]} \bm{e}^C =0,
\end{align}
suggesting that both commutators $[\ell, e_A]$ and $[e_A,e_B]$ lie in the vertical subspace. Similarly, using the definition \eqref{d-k}, it then follows that,
\begin{align}
\varphi_A = \iota_{[\ell, e_A]} \bm{k}, \qquad \text{and} \qquad w_{AB} = \iota_{[e_A, e_B]} \bm{k}.
\end{align}
All these conditions therefore determines the commutation relations of the frame fields\footnote{Our definition of the Carrollian vorticity $w_{AB}$ differs from \cite{Ciambelli:2018xat, Petkou:2022bmz} by a factor of 2.}, 
\begin{align}
\boxed{
[e_A,e_B] = w_{AB}\ell,
\qquad \text{and} \qquad
[\ell,e_A]= \varphi_A \ell. \lb{C-comm}
}
\end{align}
We comment here that the Jacobi identity of the commutators determines the evolution of the Carrollian vorticity,
\begin{align}
\ell[w_{AB}]= e_A[\varphi_B] -e_B[\varphi_A]. 
\end{align}

It is important to appreciate that, as we have already derived, the commutator between horizontal basis vectors $[e_A,e_B]$ does not lie in the horizontal subspace $\text{\bf hor}(H)$ when the Carrollian vorticity $w_{AB}$ does not vanish. Geometrically speaking, following from the Frobenius theorem, this means that the horizontal subspace $\text{\bf hor}(H)$ is not integrable in general, meaning that it cannot be treated as a tangent space to a $2$--dimensional submanifold of the space $H$.

Given the metric $q_{AB}$ on the sphere $S$, we define the \emph{expansion tensor} $\theta_{AB}$ as the change of the sphere metric along the vertical direction,
\begin{align}
\theta_{AB}  := \frac{1}{2} \cL_\ell q_{AB} = \frac{1}{2}\ell[q_{AB}].
\end{align}
The trace of the expansion tensor, called the \emph{expansion} and denoted by $\theta$, computes the change of the area element of the sphere $S$ along the vector $\ell$,
\begin{align}
 \theta := q^{AB} \theta_{AB} = \ell[\ln \sqrt{q}].
\end{align}


\subsection{Horizontal Covariant Derivative}

Another ingredient that is needed in order to write the Carrollian conservation laws is the notion of the horizontal covariant derivative. To this end, we introduce the Christoffel-Carroll symbols \cite{Ciambelli:2018xat} defined in the same manner as the standard Christoffel symbols but using the $2$-sphere metric and the horizontal basis vectors,
\begin{align}
{}^{\sss (2)}\Gamma^A_{BC} := \frac{1}{2} q^{AD}\left( e_B[ q_{DC}] +e_C[ q_{BD}] - e_D [q_{BC}] \right). \lb{Chris-Car}
\end{align}
It is torsion-free, ${}^{\sss (2)}\Gamma^A_{BC} = {}^{\sss (2)}\Gamma^A_{CB}$ by definition. We then define the \emph{horizontal covariant derivative} (or sometimes called the Levi-Civita-Carroll covariant derivative \cite{Ciambelli:2018xat}) $\sD_A$ which acts on a horizontal tensor $T = T^A{}_B e_A \otimes \bm{e}^B$ as 
\begin{align}
\sD_A T^B{}_C = e_A [T^B{}_C ]+ {}^{\sss (2)}\Gamma^B_{DA} T^D{}_C - {}^{\sss (2)}\Gamma^D_{CA} T^B{}_D,
\end{align}
and it can straightforwardly be generalized to a tensor of any degrees. By construction, the sphere metric $q_{AB}$ is compatible with this connection, that is $\sD_C q_{AB} =0$. 

One useful formula will be that the horizontal divergence of a horizontal vector $X = X^A e_A$ is given by
\begin{align}
\sD_A X^A = \frac{1}{\sqrt{q}} e_A \left[ \sqrt{q}X^A \right].
\end{align}
More details on this covariant derivative are provided in Appendix \ref{hor-derivative}.


\subsection{Adapted coordinates for the Carroll structure} \lb{sec-coordinate}

Up until this stage, we have always kept our presentation of the Carroll structure abstract and is thus completely independent of the choices of coordinates on the space $H$. We can pretty much continue this trend for the rest of this article. However, some physical pictures can be easily garnered when working explicitly with coordinates and, for practical purposes, some computations are conveniently carried out when expressing in coordinates. We will discuss the coordinate choices in this section.  

Since the space $H$ is structured as the fiber bundle over the sphere $S$, we can, without loss of generality, choose a general coordinate system $x^i = (u,y^A)$ such that open sets of the cuts at $u=\mathrm{constant}$, which denoted by $S_u$, are identified with open sets of the sphere $S$ through the projection map, $S_u  \to S$, which maps the coordinates $y^A$ to the coordinates on the sphere\footnote{More rigorously, $p^A$ is a transition map, $p^A := (\s \circ p \circ x^{-1} (u,y))^A$, where $x: H \to \mathbb{R}^{D-1}$ and $\s: S \to \mathbb{R}^{D-2}$ provide, respectively, local coordinates on $H$ and $S$. },
\begin{align}
y^A \to \sigma^A= p^A(u,y^B). 
\end{align}
In what follows, we will denote the Jacobian of the push-forward by $J: T S_u\to TS$, and it is explicitly given in coordinates by 
$J_A{}^B =\pa_A p^B$, where we have used the notation $\pa_A := \frac{\pa}{\pa y^A}$. In this general coordinate system, the Carroll structure is then characterized by a scale factor $\alpha$ and a velocity field $V^A$ such that
\begin{align}
\ell =  e^{-\alpha} D_u, \qquad \text{and} \qquad \bm{e}^A = (\bm{\rd} y^B - V^B\bm{\rd} u) J_B{}^A,
\end{align}
where we defined $D_u := (\pa_u + V^A \pa_A)$. Following from the definition of the co-frame field $\bm{e}^A := p^*(\bm{\rd} \s^A)$, the velocity field $V^A$ can be expressed in terms of the projection map as 
\begin{align}
V^A = - \pa_u p^B  (J^{-1})_B{}^A, \qquad \text{such that} \qquad D_u p^A = 0,
\end{align}
where we introduced the matrix $J^{-1}$ to be the inverse of the Jacobian such that $J_A{}^C (J^{-1})_C{}^B = (J^{-1})_A{}^C J_C{}^B = \delta_A^B$. Let us also remark here that a change of the scale factor $\alpha$ preserves the Carroll structure while a variation of the velocity field $V^A$ changes the Carroll structure. It follows from the definition of the Jacobian that 
\begin{align}\label{Carrollian-0}
\pa_B J_C{}^A = \pa_C J_B{}^A.
\end{align}
In addition, the property $\bm{\rd e}^A =0$ imposes the following constraint on the Carrollian velocity and the Jacobian,
\begin{align}\label{Carrollian}
D_u J_B{}^A  = -(\pa_B V^C) J_C{}^A, \qquad \text{and} \qquad D_u (J^{-1})_B{}^A  = (J^{-1})_B{}^C\pa_C V^A. 
\end{align}

The Ehresmann connection, obeying the condition $\iota_\ell \bm{k}$, is characterized by the \emph{Carrollian connection density}, $\beta_A$, and it can be parameterized as
\begin{align}
\bm{k} = \e^\alpha (\bm{\rd} u -  \beta_A \bm{e}^A).
\end{align}
The choice of the Ehresmann connection also fixes the form of the horizontal basis vectors $e_A$ by the conditions, $\iota_{e_A} \bm{k} =0$ and also $\iota_{e_A} \bm{e}^B = \delta^B_A$. In our parameterization, the horizontal basis is given by
\begin{align}
e_A = (J^{-1})_A{}^B\pa_B +\beta_A D_u.
\end{align}

In this general coordinate system, we can evaluate the Carrollian commutators and in turn obtain the coordinate expression of the Carrollian acceleration $\varphi_A$ and the Carrollian vorticity $w_{AB}$ (see Appendix \ref{Phi-W-derive}). They are given by 
\begin{align}
\varphi_A &=  D_u \beta_A +e_A[\alpha], \\
w_{AB} &= \e^\a \left( e_A[\beta_B] - e_B[\beta_A] \right).  
\end{align}

In this article, we will always work with the general coordinates $x^i = (u,y^A)$ on the space $H$ as they are, by construction, independent of the Carroll structure. Let us, however, mention that we can also choose to work with the \emph{adapted coordinates} $(u,\sigma^A)$ on $H$ which are such that the action of the projection is trivial, $p:(u,\sigma) \to \sigma$. With this choice, the coordinate $u$ is regarded as the fiber coordinate. By definition, the velocity field $V^A =0$ vanishes in the adapted coordinates. These coordinates are therefore co-moving coordinates, which are such that 
\begin{align} 
\ell = \e^{-\a}\pa_u, \qquad \text{and} \qquad  \bm{e}^A = \bm{\rd} \sigma^A.
\end{align} 
To connect with the previous parameterization, one can derive, given the coordinates $y^A(u,\sigma)$, the following relations
\begin{align}
 V^A = \frac{\pa y^A}{\pa u}, \qquad \text{and} \qquad (J^{-1})_A{}^B = \frac{\pa y^B}{\pa \sigma^A}.
\end{align}
The Ehresman connection in the adapted coordinates therefore reads 
\begin{align}
\bm{k}= \e^\alpha \left( \bm{\rd} u - \beta_A \bm{\rd} \sigma^A \right).
\end{align}
The expressions for the the Carrollian acceleration and the Carrollian vorticity simplifies in the co-moving coordinates becomes
\begin{align}
\varphi_A &=
\left( \frac{\pa}{\pa \s^A} + \beta_A \pa_u \right) \alpha + \pa_u \beta_A, \\
w_{AB} &= \e^\alpha  \left(  \left( \frac{\pa}{\pa \s^A} + \beta_A \pa_u \right) \beta_B  - \left( \frac{\pa}{\pa \s^B} + \beta_B \pa_u \right) \beta_A \right). 
\end{align}
The co-moving coordinates have been widely adopted in the Carrollian literature (see for example \cite{Duval:2014uoa, Ciambelli:2019lap, Ciambelli:2018ojf}) as the apparent absence of the velocity field and the Jacobian factor heavily simplifies all computations. Also, this choice of coordinates works well when considering field variations that leave the Carroll structure unchanged. We will, however, be more general by considering the set of variations that can change the Carroll structure, and will therefore work with the general, field-independent, coordinates $x^i = (u,y^A)$. 

Let us also comment that the vorticity is the curvature of the Witt connection 
\be
w_{AB} = \e^\alpha\left(\frac{\pa}{\pa \s^A} \b_B -\frac{\pa}{\pa \s^B} \b_A + [\b_A,\b_B]_{\mathrm{W}}\right),
\qquad \text{where} \qquad
[a,b]_{\mathrm{W}} := a\pa_u b-b\pa_ua.
\ee
The bracket $[\ ,\ ]_{\mathrm{W}}$ is the Witt bracket\footnote{This means that $\beta_A \pa_u$ is an element of the Witt algebra. Let us also comment that it is more common to work with the Laurent polynomial basis $L_n := - u^{n+1} \pa_u$ where now $\beta_A(u,\s) = \sum\limits_{n \in \mathbb{Z}} \beta_A^{(n)}(\s) L_n$. In this basis, the Witt algebra is in the well-familiar form, $[L_n,L_m]_{\text{W}} = (n-m)L_{n+m}$.}. 
This means that the corresponding symmetry group is the group $\mathrm{Diff}(\mathbb{R})$ of a space-dependent time reparameterizations. 
An element of this group is denoted $\hat{U}$ and simply represented by a function $\hat{U}:u\to U(u, \s) $. 
The demand that the vorticity vanishes $w_{AB}=0$  means that $\beta_A \pa_u =  -\hat{U}^{-1} \circ \pa_A \hat{U}$ is a flat $\mathrm{Diff}(\mathbb{R})$ connections. This implies that the coefficient $\beta_A$ is given\footnote{ This follows from the fact that $[\hat{U}^{-1} \phi](u,\sigma): = \phi(U(u,\sigma),\sigma)$ which  gives 
\bea
[\pa_A \hat{U}^{-1}] \phi &=& [\pa_AU] [\pa_u\phi](U,\sigma),\cr
[ \hat\beta_A \circ \hat{U}^{-1} ] \phi(u,\sigma) &=& \beta_A \pa_u \phi( U,\sigma) = [\beta_A \pa_uU] (\pa_u\phi ) (U,\sigma),
\eea
where we denoted $\hat{\beta}_A :=\beta_A \pa_u$. 
}, in a comoving coordinate system, by 
\be
\b_A = -\frac{\pa_A U}{\pa_u U}.
\ee
The slices $U= \text{constant}$ are then the Bondi slices.
In an arbitrary coordinate system $\beta_A$ can be written simply as $\beta_A = e_A[u + U]$. 


\subsection{Carrollian transformations} 

We conclude our geometrical setup on Carroll structures by discussing Carrollian diffeomorphism. In general, there are two types of diffeomorphism of the space $H$ --- one that preserves the fiber bundle structure and one that changes it. Here we will focus on the former case and we will discuss the latter case when considering hydrodynamics in the next section. 

Transformations that preserve the fiber bundle structure of the space $H$, which has been particularly referred to as \emph{Carrollian transformations} or \emph{Carrollian diffeomorphism} in the literature, are such that 
\begin{align}
u \to u'(u,\s^A), \qquad \text{and} \qquad \s^A \to \s'^A(\s^B). 
\end{align}
In this class of transformations, the co-frame field $\bm{e}^A$ only changes by the diffeomorphism on the sphere $S$, inferring that the basis vector $\ell$ can only change by rescaling, $\delta^\Carr \ell \propto \ell$. In other words, the new Carrollian vector still belongs to the equivalence class $[\ell]_{\sim}$. This therefore means that the velocity field is unchanged under Carrollian transformations,
\begin{align}
\delta^\Carr V^A = 0.
\end{align} 
We now compute how the components $(\alpha, \beta_A, q_{AB})$ of the Carroll structure vary under infinitesimal Carrollian diffeomorphism generated by a vector field
\begin{align}
\xi = \tau \ell + Y^A e_A, \qquad \text{where} \qquad \ell[Y^A] =0,
\end{align} 
and $\tau$ is a generic function on the space $H$. It follows from
\begin{align}
\delta_\xi \ell = \cL_\xi \ell = [\xi, \ell] = -\left( \ell[\tau] + Y^A \varphi_A \right) \ell,
\end{align} 
and $\delta^\Carr \ell = -(\delta^\Carr \alpha)\ell$. So the transformation of the scale factor is 
\begin{align}
\delta^\Carr_{(\tau,Y)} \alpha = \ell[\tau] + Y^A \varphi_A.  \lb{del-C-a}
\end{align}
For the Carrollian connection $\beta_A$, we use that $\delta^\Carr_{(\tau,Y)} \bm{k} = \cL_\xi \bm{k}$ to read off the transformation of $\beta_A$, which is 
\begin{equation}
\begin{aligned}
- \e^\a \bbdelta^\Carr_{(\tau,Y)} \beta_A   =  (e_A - \varphi_A)[\tau] + w_{AB} Y^B,
\lb{del-C-b}
\end{aligned}
\end{equation}
where we defined the variation $\bbdelta^\Carr \beta_A :=(J^{-1})_A{}^B \delta^\Carr(J_B{}^C \beta_C)$. Lastly, we use that $\delta^\Carr_{(\tau,Y)} q =  \cL_\xi q$ to show that the sphere metric $q_{AB}$ transforms as
\begin{align}
\bbdelta^\Carr_{(\tau,Y)} q_{AB} = 2\left( \tau \theta_{AB} + \sD_{(A} Y_{B)} \right),\lb{del-C-q}
\end{align}
where we defined $\bbdelta^\Carr q_{AB} := (J^{-1})_A{}^C (J^{-1})_B{}^D \delta^\Carr (J_C{}^E J_D{}^F q_{EF})$. Let us also note that one can consider Carrollian isometries such that $\cL_\xi q = 0$ or conformal Carrollian isometries such that $\cL_\xi q = \Omega q$, for a conformal factor $\Omega$. In such cases, we will have more constraints on the transformation parameters $(\tau, Y)$ (see for instance the discussions in \cite{Ciambelli:2019lap, Ciambelli:2018ojf, Donnay:2019jiz, Duval:2014uva}). 

\section{Carrollian Hydrodynamics} \lb{hydro}

Having formally established essential elements of the Carroll structure, we proceed to the discussion of hydrodynamics and its ultra-relativistic cousin, namely the \emph{Carrollian hydrodynamics}. It has been well established fact that Galilean fluids can be derived by taking the non-relativistic limit, $c \to \infty$, of the general relativistic energy-momentum tensor $T^{ij}$ and their corresponding dynamics are therefore controlled by the non-relativistic version of the conservation laws, $\nabla_j T_i{}^j =0$. The equations governing the `Galilean' time evolution of the fluid are the continuity equation, energy conservation equation, and the Navier-Stokes equations. In a much similar spirit, taking the Carrollian, $c \to 0$, leads to a new, and peculiar, kind of fluids and their corresponding hydrodynamic equations that are Carrollian-covariant \cite{Ciambelli:2018xat}. In this section, we will present how the Carrollian hydrodynamic equations can be obtained from the $c \to 0$ contraction of the relativistic conservation laws.

\subsection{Metric on $H$}

Until this stage, the geometry of the space $H$ have been constructed from the Carroll structure which relied on the concept of fiber bundle. In order to discuss the conservation equations of the fluid energy-momentum tensor, $\nabla_j T_i{}^j =0$, the space $H$ needs to be equipped with an additional structure: a $3$--dimensional Lorentzian metric $h = h_{ij} \bm{\rd} x^i \otimes \bm{\rd} x^j$ and the Levi-Civita connection $\nabla$ compatible with it. We will discuss the metric first. 

We are considering a family of Lorentzian matrices whose elements are labelled by a single real parameter, the \emph{speed of light}\footnote{In practice, it is the square of the speed of light, $c^2$, that will enter the computations.} $c$ and constructed entirely from the data of the Carroll geometry discussed in the previous section. By doing so, we ensure that the chosen metric is covariant under Carrollian diffeomorphism. We further make the following assumptions on the components of the metric\footnote{The second condition $h(\ell, e_A) =0$, in fact, can be relaxed by choosing $h(\ell, e_A) = c^2 \e^\a B_A$ for an arbitrary function $B_A$. The choice of $B_A$ is gauge as one can always absorb $B_A$ into the definition of the horizontal basis $e_A$, and correspondingly redefine the Ehresmann connection $\bm{k}$ and the sphere metric $q_{AB}$, by shifting the Carrollian connection $\beta_A \to \beta_A + B_A$. This new basis $e'_A = e_A + B_A D_u$ then satisfies the second condition $h(\ell,e'_A) =0$.}, 
\begin{align}
h(\ell, \ell) = - c^2, \qquad h(\ell, e_A) = 0, \qquad \text{and} \qquad h(e_A, e_B) = q_{AB}. 
\end{align}
These conditions also infer that, when taking the limit $c \to 0$, the resulting metric on $H$ coincides with the null Carrollian metric, i.e., $h \stackrel{ c \to 0}{=} q$. Observe that the Carrollian vector field $\ell$ is timelike in general and becomes null in the Carrollian limit, $h(\ell, \ell) \stackrel{ c \to 0}{=} 0$. The metric $h$ and its inverse $h^{-1}$ are given in the Carrollian basis by\footnote{We use $\circ$ to denote the symmetric tensor product of tensors, i.e., $A \circ B = \frac{1}{2} \left(A \otimes B + B \otimes A \right)$}. 
\begin{equation}
\begin{aligned}
h = - c^2 \bm{k} \circ \bm{k} + q_{AB}\bm{e}^A \circ \bm{e}^B, \qquad \text{and} \qquad h^{-1} = - c^{-2} \ell \circ \ell + q^{AB}  e_A \circ e_B. \lb{RPmetric}
\end{aligned}  
\end{equation}
The inverse metric is thus singular in the Carrollian limit $c \to 0$. This particular form of the metric is known as the \emph{Randers-Papapetrou metric} and it has been utilized extensively in Carrollian physics literatures \cite{Ciambelli:2018xat, Ciambelli:2018wre,Ciambelli:2018ojf, Campoleoni:2018ltl, Petkou:2022bmz}. Also, having the metric $h$, one can derive the relations between the basis vectors and 1-forms, which are
\begin{align}
\bm{k} = -\frac{1}{c^2} h(\ell, \cdot), \qquad \text{and} \qquad \bm{e}^A = q^{AB} h(e_B, \cdot). 
\end{align}

It is important to appreciate that the metric \eqref{RPmetric} can be viewed as the expansion in the small parameter $c^2$ around the Carrollian point, $c^2 =0$. With this in mind, we will also make another assumption that the sphere metric $q_{AB}$ admits the expansion in the small parameter $c^2$ such that 
\begin{align}
q_{AB} = \mrq_{AB} + 2c^2 \lam_{AB} + \O(c^4), \qquad \text{and} \qquad q^{AB} = \mrq^{AB} - 2c^2 \lam^{AB} +\O(c^4), \lb{metric-c-expand}
\end{align}
where $\mrq^{AB}$ is the inverse of $\mrq_{AB}$ and we defined $\lam^{AB} := \mrq^{AC}\mrq^{BD} \lam_{CD}$ and $\lam := \mrq^{AB} \lam_{AB}$. Note also that, to properly manipulate the $c^2$-expansion, we will use the leading-order sphere metric $\mrq_{AB}$ and its inverse $\mrq^{AB}$ to lower and raise indices of horizontal tensors. Remarks are in order here: 

$i)$ At first glance, doing this $c^2$-expansion of the sphere metric may seem like we have introduced unnecessary complications to the problems. We will later demonstrate that this expansion is necessary to derive the hydrodynamic conservation equations from symmetries. 

$ii)$ In our derivations, it is sufficient to expand the Lorentzian metric $h$ to the order $c^2$. Therefore, we can assume that the components $\alpha$ and $\beta_A$ do not admit this $c^2$-expansion.

Since we now have the $c^2$-expansion of the sphere metric, some objects will also inherit this similar expansion. The obvious ones are the expansion tensor and its trace, which exhibit the following expansion 
\begin{align}
\theta_{AB} = \mr{\theta}_{AB} + c^2 \ell[\lam_{AB}] + \O(c^4), \qquad \text{and} \qquad \theta = \mr{\theta} + c^2 \ell [\lam] + \O(c^4),
\end{align}
where the zeroth-order terms are
\begin{align}
\mr{\theta}_{AB} = \frac{1}{2} \ell\left[\mrq_{AB}\right], \qquad \text{and}\qquad \mr{\theta} = \mrq^{AB} \mr{\theta}_{AB} = \ell\left[\ln \sqrt{\mrq}\right].
\end{align} 
Another object that will admits the $c^2$-expansion is the Christoffel-Carroll synbols $\two\Gamma^A_{BC}$, and we present its expansion in Appendix \ref{hor-derivative}. 

In order to do integration on the space $H$, we need the volume form on the $H$. We define the volume form as 
\begin{equation}
\bm{\epsilon}_H :=  \bm{k} \wedge\bm{\epsilon}_S, \qquad \bm{\epsilon}_S =  \sqrt{q}  \left( \frac{ \varepsilon_{AB} }{2} \bm{\rd} \s^A \wedge \bm{\rd} \s^B\right),
\end{equation}
where $\varepsilon_{AB}$ is the standard Levi-Civita symbol (satisfying $\varepsilon_{AC} \varepsilon^{CB} = \delta_A^B$). $\bm{\epsilon}_S$ denotes the canonical volume form on the sphere $S$, which satisfies the relation $
\iota_\ell \bm{\epsilon}_H  = p^*(\bm{\epsilon}_S)$. 
As before, using that $\sqrt{q} = \sqrt{\mr{q}}(1 + c^2 \lam) + \O(c^4)$, we thus obtain the $c^2$-expansion of the volume form,
\begin{align}
\bm{\epsilon}_H  = \left( 1+c^2\lam \right) \mr{\bm{\epsilon}}_H + \O(c^4), \qquad \text{and} \qquad \bm{\epsilon}_S  = \left( 1+c^2\lam \right) \mr{\bm{\epsilon}}_S + \O(c^4), 
\end{align}
where $\mr{\bm{\epsilon}}_H$ and $\mr{\bm{\epsilon}}_S$ denote the zeroth-order of the volume form on $H$ and on $S$, repectively.

\subsection{Covariant derivative}

Before considering Carrollian hydrodynamics, let us now consider the Levi-Civita connection $\nabla$ compatible with the metric \eqref{RPmetric}, that is $\nabla_i h_{jk} =0$. Let us compute the covariant derivative the basis vector fields, namely $\nabla_{\ell} \ell, \nabla_{e_A} \ell, \nabla_{\ell} e_A$, and $\nabla_{e_A} e_B$, as they will become handy tools when evaluating the hydrodynamic conservation equations. We start with the covariant derivative $\nabla_{\ell} \ell$, which we will present the computation in full detail here. Complete derivations of the others, which are done in a similar vein, are provided for the readers in Appendix \ref{cov-deri}. The term $\nabla_\ell \ell$, can be decomposed as
\begin{align}
\nabla_{\ell} \ell = (k_i \nabla_{\ell} \ell^i) \ell + (q^{AB}e_{Bi} \nabla_{\ell} \ell^i)e_A.
\end{align}
Using the metric $h$ and the Leibniz rule, one can show that the vertical component vanishes\footnote{This correspond to a choice of vanishing inafinity  
$
\kappa =  \ell[\ln c]=0.
$ } as follows:
\begin{equation}
\begin{aligned}
k_i \nabla_{\ell} \ell^i = - \frac{1}{c^2} h \left(\ell, \nabla_{\ell} \ell \right) =  - \frac{1}{2c^2} \ell \left[ h\left(\ell, \ell \right) \right] =  0,
\end{aligned}
\end{equation}
as $h(\ell, \ell) = -c^2$ is constant. The horizontal components can be evaluated with the help of the commutation relations \eqref{C-comm} as follows:
\begin{equation}
\begin{aligned}
e_{Bi} \nabla_{\ell} \ell^i = h\left( e_B, \nabla_{\ell} \ell \right) &= -h\left(  \ell , \nabla_{\ell}e_B \right) \\
& =  -h\left(  \ell , [\ell,e_B] \right) - \frac{1}{2} e_B[ h \left(\ell, \ell \right)] \\
& =  c^2 \varphi_B. 
\end{aligned}
\end{equation}
Therefore, the covariant derivative of the vertical vector field along itself is given by 
\begin{align}
\nabla_{\ell} \ell = c^2 \varphi^A e_A + \O(c^4).
\end{align}
Observe that it vanishes in the Carrollian limit $c^2 \to 0$, dictating that the vector $\ell$ is the null generator of null geodesics on the space $H$. 

The covariant derivative of the vertical vector along the horizontal vectors can be computed using the same technique. One can show that (see Appendix \ref{cov-deri}) it is given by 
One could more simply write
\begin{align}
\nabla_{e_A} \ell= \left(\mr{\theta}_A{}^B + c^2 \left(\frac{1}{2}w_A{}^B + \ell[\lam_{A}{}^B]  \right)\right)e_B + \O(c^4).
\end{align}
where $ \lam_{A}{}^B = \mrq^{BC} \lambda_{AC}$. The covariant derivative of the horizontal basis along the vertical basis, $\nabla_{\ell} e_A$, is already determined from $\nabla_{e_A} \ell$ and the commutator $[\ell, e_A]$. We are left with the remaining covariant derivative, $\nabla_{e_A} e_B$. Its vertical component, $k_i \nabla_{e_A} e_B{}^i$ can be inferred from $\nabla_{e_A} \ell$. For the horizontal components, $e^C{}_i \nabla_{e_A} e_B{}^i$, using that $q_{AB} = h \left(e_A, e_B \right)$ and the definition of the Christoffel-Carroll symbols \eqref{Chris-Car}, we can show that 
\begin{equation}
\begin{aligned}
\nabla_{e_A} e_B = \ & \left(\frac{1}{c^2} \mr{\theta}_{AB} + \left(\frac{1}{2}w_{AB} + \ell[\lam_{AB}]\right) \right) \ell + \two\mr{\Gamma}^C_{AB} e_C \\
&+c^2 \left( \sD_A\lam_B{}^C +\sD_B\lam_A{}^C-\sD^C\lam_{AB} \right)e_C.
\end{aligned}
\end{equation}

With all these results, one can calculate the spacetime divergence of the basis vectors. Using the decomposition \eqref{C-projector}, we obtain
\begin{equation}
\begin{aligned}
\nabla_i \ell^i = \delta_i{}^j \nabla_j \ell^i &= \left( k_i \ell^j + e^B{}_i e_B{}^j \right) \nabla_j \ell^i  = \mr{\theta}+c^2 \ell[\lam],
\end{aligned} 
\end{equation}
and in a similar manner, 
\begin{equation}
\begin{aligned}
\nabla_i e_A{}^i = \delta_i^j \nabla_j e_A{}^i &= \left( k_i \ell^j + e^B{}_i e_B{}^j \right) \nabla_j e_A{}^i  =  \varphi_A + \two\mr{\Gamma}^B_{AB} + c^2 e_A [\lam] + \O(c^4). 
\end{aligned} 
\end{equation}

It is important to remark that the 3-dimensional metric compatible connection $\nabla_i$ contains a component that diverges when taking the Carrollian limit $c \to 0$. This is to be expected since the inverse metric \eqref{RPmetric} diverges in this limit. This also suggests that, in practical, computations have to be carried out at finite value of $c$ and the Carrollian limit needs to be taken at the last step. 

\subsection{Carrollian Hydrodynamics}

Armed with all these tools, we are ready to discuss the hydrodynamics of Carrollian fluid. Let us start from the general form of relativistic energy-momentum tensors, 
\begin{align}
T^{ij} = (\sE+ \sP) \frac{\ell^i \ell^j}{c^2} + \sP h^{ij} + \frac{q^i \ell^j}{c^2} + \frac{q^j \ell^i}{c^2} + \tau^{ij}, \lb{rela-em}
\end{align}
where we chose the vertical vector $\ell$ to be the fluid velocity. The variables appeared in the fluid energy-momentum tensor consist of the fluid internal energy density $\sE$, the fluid pressure $\sP$, the heat current $q^i$, and the viscous stress tensor $\tau^{ij}$, which is symmetric and traceless. The latter two quantities represent dissipative effects of the fluid and, by construction, they obey the orthogonality conditions with the fluid velocity, $q_i \ell^i =0$ and $ \tau_{ij}\ell^j=0$. This means that, in light of Carrollian geometry we have introduced, these dissipative tensors are horizontal tensors,
\begin{align}
q^i = q^A e_A{}^i, \qquad \text{and} \qquad \tau^{ij} = \tau^{AB} e_A{}^i e_B{}^j.
\end{align}

We are interested in the mixed indices version of the fluid energy-momentum tensor. Using the metric \eqref{RPmetric}, it is given by 
\begin{align}
T_i{}^j &= - \left( \sE \ell^j + q^A e_A{}^j\right) k_i + \left(\frac{1}{c^2} q_{AB}q^B  \ell^j + \left(q
_{AC}\tau^{CB} + \sP \delta_A^B\right)  e_B{}^j \right) e^A{}_i. 
\end{align}
Furthermore, we choose the following $c^2$-dependence \cite{Ciambelli:2018wre, Ciambelli:2018xat, Ciambelli:2019kiw} of the dissipative tensors,
\begin{align}
q^A = \sJ^A + c^2\left( \pi^A  - 2\lam^A{}_B \sJ^B \right), \qquad \tau^{AB} = \frac{\Sigma^{AB}}{c^2} + \sS^{AB}.  \lb{q-tau}
\end{align}
Note also that $q_{AB}q^B =\sJ_A + c^2 \pi_A + \O(c^4)$. Following from this parameterization, the fluid energy-momentum tensor can be expressed as the expansion in $c^2$ as 
\begin{align}
T_i{}^j = \frac{1}{c^2} T^{\sss(-1)}_i{}^j +T^{\sss(0)}_i{}^j +\O(c^2), \lb{T-fluid}
\end{align}
where each term reads
\begin{subequations}
\begin{align}
T^{\sss(-1)}_i{}^j &= \left(\sJ_A \ell^j +\Sigma_A{}^B e_B{}^j \right) e^A{}_i \\
T^{\sss(0)}_i{}^j &= - \left( \sE \ell^j + \sJ^A e_A{}^j\right) k_i + \left( \pi_A \ell^j + \left( \sK_A{}^B + \sP \delta_A{}^B \right)e_B{}^j\right)e^A{}_i, \lb{e-m-fluid}
\end{align}
\end{subequations}
and we defined for convenience the combination,
\begin{align}
\sK_A{}^B :=  \sS_A{}^B + 2\lam_{AC} \Sigma^{CB}.
\end{align}

The dynamics of the relativistic fluid is governed by the relativistic conservation laws, $\nabla_j T_i{}^j$. Let us first evaluate the vertical component of the conservation equations. With all the tools we derived previously, we show that 
\begin{equation}
\begin{aligned}
\ell^i \nabla_j T_i{}^j &= \nabla_j \left(\ell^i T_i{}^j\right) - T_i{}^j \nabla_j \ell^i \\
&= -\nabla_j \left( \sE \ell^j +q^A e_A{}^j  \right) - \frac{1}{c^2}q^A \left(e_{Ai} \nabla_{\ell} \ell^i \right) - \left(\tau^{AB} + \sP q^{AB}\right) \left( e_{Ai} \nabla_{e_B} \ell^i\right) \\
& =- (\ell + \theta )[\sE] - \sP \theta - (\sD_A + 2 \varphi_A) q^A - \tau^{AB} \theta_{AB} \\
& = \frac{1}{c^2}\mathbb{C}+  \mathbb{E}  + \O(c^2),
\end{aligned}
\end{equation}
where the coefficients of the $c^2$-expansion are 
\begin{align}
\mathbb{E}  &= -(\ell+ \mr{\theta})[\sE] - \sP \mr{\theta} - (\mrD_A + 2 \varphi_A) \sJ^A   -\sS^{AB} \mr{\theta}_{AB} - \Sigma^{AB} \ell[\lam_{AB}], \lb{E-eq}  \\
\mathbb{C} & = -\Sigma^{AB} \mr{\theta}_{AB}. \lb{C-eq}
\end{align}
Imposing $\ell^i \nabla_j T_i{}^j =0$ as one taking the limit $c \to 0$ demands $\mathbb{E} =0$ and $\mathbb{C} =0$. The first equation is the Carrollian energy evolution equation and second equation is the constraint equation. Note that the expression $\mathbb{E}$ for the energy equation differs from the original work \cite{Ciambelli:2018xat} due to the presence of the tensor $\lam_{AB}$ and the fluid velocity $V^A$ contained implicitly in the Carrollian $\ell$. As we will discuss in the next section, these two additional variables are part of the phase space of Carrollian fluids and they are necessary when one wants to derive Carrollian conservation laws from symmetries. In this sense, our results generalizes those presented in \cite{Ciambelli:2018xat}.

In a similar manner to the vertical component, we compute the horizontal components of the conservation laws and consider the $c^2$-expansion. This is given by 
\begin{equation}
\begin{aligned}
e_A{}^i \nabla_j T_i{}^j &= \nabla_j \left(e_A{}^i T_i{}^j\right) - T_i{}^j \nabla_j e_A{}^i \\
&= \nabla_j \left(\frac{1}{c^2} q_{AB}q^B  \ell^j + \left(q_{AC}\tau^{CB} + \sP \delta_A^B\right)  e_B{}^j \right) + \left( \sE k_i - \frac{1}{c^2}q^B e_{Bi} \right)\nabla_{\ell}e_A{}^i\\
& \ \ \ \ + \left( q^Bk_i - \left( q_{CD}\tau^{BD} +\sP \delta^B_C \right)e^C{}_i \right)\nabla_{e_B}e_A{}^i \\
&= \frac{1}{c^2}(\ell+ \theta)[q_{AB} q^B]+ \sE \varphi_A - w_{AB} q^B + (\sD_B + \varphi_B)(q_{AC} \tau^{CB}+ \sP \delta_A^B) \\
& = \frac{1}{c^2}\mathbb{J}_A+ \mathbb{P}_A  + \O(c^2),
\end{aligned}
\end{equation}
where the zeroth-order term is
\begin{equation}
\begin{aligned}
\mathbb{P}_A = \ & (\ell+ \mr{\theta})[\pi_A]+ \sE \varphi_A - w_{AB} \sJ^B + (\mrD_B + \varphi_B)(\sK_A{}^B  + \sP \delta_A^B)  \\
&  + \left( \ell[\lam]\sJ_A + \Sigma_A{}^B \mrD_B \lam + \Sigma^{BC} \mrD_A \lam_{BC} \right),  \lb{P-eq}
\end{aligned}
\end{equation}
while the other term is
\begin{align}
\mathbb{J}_A &= (\ell + \mr{\theta})[\sJ_A ] + (\mrD_B + \varphi_B) \Sigma^B{}_A \lb{J-eq}. 
\end{align}
Taking the Carrollian limit $c \to 0$ of the conservation laws, $e_A{}^i \nabla_j T_i{}^j = 0$, imposes the Carrollian momentum evolution, $\mathbb{P}_A =0$ and the conservation of Carrollian current, $\mathbb{J}_A =0$. Again, our expression for $\mathbb{P}_A$ is the generalization of \cite{Ciambelli:2018xat}. 

Let us comment here the case when the sub-leading components of the sphere metric vanishes, $\lam_{AB} =0$ simplifies the Carrollian evolution equations,
\begin{subequations}
\begin{align}
\mathbb{E}  &= -(\ell+ \mr{\theta})[\sE] - \sP \mr{\theta} - (\mrD_A + 2 \varphi_A) \sJ^A   -\sS^{AB} \mr{\theta}_{AB},   \\
\mathbb{P}_A &=  (\ell+ \mr{\theta})[\pi_A]+ \sE \varphi_A - w_{AB} \sJ^B + (\mrD_B + \varphi_B)(\sS_A{}^B + \sP \delta_A^B),  \\
\mathbb{J}_A &= (\ell +\mr{\theta})[\sJ_A ] + (\mrD_B + \varphi_B) \Sigma^B{}_A, \\
\mathbb{C} & = -\Sigma^{AB} \mr{\theta}_{AB}.
\end{align}
\end{subequations} 
These are the Carrollian fluid equations given in the literature \cite{Ciambelli:2018xat, Ciambelli:2018ojf}. 
Note that the solutions of these equations are invariant under the shift 
$( \sE,  \pi_A ,  \sS_A{}^B ) \to ( \sE,  \pi_A + a \sJ_A,  \sS_A{}^B  + a \Sigma_A{}^B)$ where $a$ is an arbitrary parameter and where $(\sP,\sJ_A, \Sigma_{A}{}^B)$ is unchanged.

\section{Hydrodynamics from Symmetries} \lb{hydro-sym}

In this section, we tackle the Carrollian hydrodynamics from a different, but nonetheless equivalent, perspective. Our objective is to re-derive the equations that govern Carrollian hydrodynamics \eqref{E-eq}, \eqref{C-eq}, \eqref{P-eq}, and \eqref{J-eq} from the symmetries of the space $H$. 

\subsection{The Action for Carrollian Fluid}
Since the metric $h$ is defined on the space $H$, we can consider the action of the fluid whose variation yields the fluid energy-momentum tensor. We will consider the fluid action that is finite when taking the Carrollian limit $c \to 0$. The variation of the fluid action we will use takes the form 
\begin{tcolorbox}[colback = white]
\vspace{-15pt}
\begin{equation}
\begin{aligned}
\delta S_{\text{fluid}} = -\int_H \left( \sE \bbdelta \alpha  - \e^\a\sJ^A \bbdelta \beta_A  + \e^{-\a} \tilde{\pi}_A \bbdelta V^A - \frac{1}{2}\left( \tilde{\sS}^{AB} +\sP \mrq^{AB}\right) \bbdelta \mrq_{AB} - \Sigma^{AB} \bbdelta \lam_{AB}\right)\mr{\bm{\epsilon}}_H \lb{S-fluid}.
\end{aligned}
\end{equation}
\end{tcolorbox}
We defined the momentum conjugated to the velocity field $V^A$ and the leading-order sphere metric $\mrq_{AB}$ to be
\begin{align}
\tilde{\pi}_A &:= \pi_A + \lam\sJ_A  \\
\tilde{\sS}^{AB} &:= \sS^{AB} +\lam \Sigma^{AB}. 
\end{align}
We also absorbed the Jacobian factors and the velocity field variation into the definition of the variation $\bbdelta$ as follows,
\begin{align}
\bbdelta\a &:= \delta \a + \beta_A \bbdelta V^A, \\
\bbdelta\beta_A & :=  (J^{-1})_A{}^C\delta \left( J_C{}^B\beta_B \right) - (\beta \cdot \bbdelta V)\beta_A, \\
\bbdelta \mrq_{AB} &:= (J^{-1})_A{}^C (J^{-1})_B{}^D\delta \left(J_C{}^E J_D{}^F \mrq_{EF}\right) - 2\mrq_{C(A} \beta_{B)} \bbdelta V^C, \\
\bbdelta \lam_{AB} &:= (J^{-1})_A{}^C (J^{-1})_B{}^D\delta \left(J_C{}^E J_D{}^F \lam_{EF}\right) - 2\lam_{C(A} \beta_{B)} \bbdelta V^C,
\end{align}
and that we define
\begin{align}
\bbdelta V^A &:= \left(\delta V^B\right)J_B{}^A. 
\end{align}

The action \eqref{S-fluid} is simply derived from the fluid energy-momentum tensor $T^{ij}$ and the metric variation $\delta h_{ij}$. To see this, let us consider an action $S[h_{ij}]$ and its metric variation yields the energy-momentum tensor,
\begin{align}
\delta S = \int_H \left( \frac{1}{2} T^{ij} \delta h_{ij} \right)\bm{\epsilon}_H
\end{align}
Since the fluid energy-momentum tensor \eqref{T-fluid} has a part that diverges when taking the Carrollian limit $c \to 0$, the variation $\delta S$ also diverges in this limit. To obtain the finite action \eqref{S-fluid}, we subtract the divergent part from $\delta S$ then take the Carrollian limit, that is
\begin{align}
\delta S_{\text{fluid}} := \lim_{c \to 0} \left( \delta S - \frac{1}{c^2}\delta S_{\sss (-1)}\right).
\end{align}
We note that the divergent part is given by 
\begin{align}
\delta S_{\sss (-1)} := \lim_{c \to 0} \left(c^2 \delta S \right) = \int_H \left( \frac{1}{2} T_{\sss (-1)}^{ij} \delta h_{\sss(0)}{}_{ij} \right)\mr{\bm{\epsilon}}_H,
\end{align}
where we used that the metric variation is regular as $c \to 0$ and schematically expands as $\delta h_{ij} = \delta h_{\sss(0)}{}_{ij} + c^2 \delta h_{\sss(1)}{}_{ij} + \O(c^4)$. The fluid action \eqref{S-fluid} is thus
\begin{align}
\delta S_{\text{fluid}} = \int_H \frac{1}{2}\left( T^{\sss (0)}{}^{ij} \delta h_{\sss(0)}{}_{ij} +T^{\sss (-1)}{}^{ij} \delta h_{\sss(1)}{}_{ij} + \lam T^{\sss (-1)}{}^{ij} \delta h_{\sss(0)}{}_{ij}\right)\mr{\bm{\epsilon}}_H. \label{var}
\end{align}

\subsection{Near-Carrollian Diffeomorphism} \lb{near-ultra}

To derive the Carrollian hydrodynamic equations from the variation of the action \eqref{S-fluid} under certain symmetries, we first need to specify those symmetries and derive the symmetry transformations for the metric components, $(\alpha, \beta_A, V^A, \mrq_{AB}, \lam_{AB})$. The seemingly obvious choice one could consider is the Carrollian diffeomorphism. However, Carrollian diffeomorphism is not sufficient to derive the complete set of hydrodynamic equations \eqref{E-eq}, \eqref{C-eq}, \eqref{P-eq}, and \eqref{J-eq}, as already shown in \cite{Ciambelli:2018ojf}. The reasons for this limitation are as follows:

$i)$ Carrollian diffeomorphism fixes the variation of the velocity field, $\delta^\Carr V^A =0$, hence turning off a phase space degree of freedom conjugated to the velocity, that is the fluid momentum density. 

$ii)$ There are only two symmetry parameters $(\tau, Y^A)$ for the Carrollian diffeomorphism, while there are four hydrodynamic equations. The symmetries labelled by the parameter $\tau$ and $Y^A$ correspond, respectively, to the the energy equation \eqref{E-eq} and the momentum equation \eqref{P-eq}.
To obtain the remaining two equations, the current conservation \eqref{J-eq} and the constraint \eqref{C-eq}, we need two more symmetry parameters.

\noindent We therefore need to detach our consideration from the Carrollian diffeomorphism and consider a general diffeomorphism on the space $H$. The general diffeomorphism on $H$ is labelled by vector fields of the form,
\begin{align}
\xi = f \ell + X^A e_A,
\end{align}
where $f$ and $X^A$ are arbitrary functions on $H$. This general diffeomorphism will definitely change the Carroll structure. In the same fashion as our prior discussions, let us expand the transformation parameters $(f, X^A)$ in the small parameter $c^2$ as
\begin{align}
f = \tau + c^2 \psi + \mathcal{O}(c^4), \qquad \text{and} \qquad X^A = Y^A + c^2 Z^A+\mathcal{O}(c^4), \lb{fluid-diff}
\end{align}
where now the parameter $(\tau,\psi,Y^A,Z^A)$ are functions on $H$. This way, we have already secured four parameters we need for four equations of Carrollian fluid. It is of extreme importance to point out that expanding the diffeomorphism around $c^2 =0$ can be regarded as the analog to the diffeomorphism of spacetime geometry in the close vicinity of a black hole horizon, the near-horizon diffeomorphism, with $c^2$ plays the same role as the distance away from the black hole horizon. We will refer to this diffeomorphism as the \emph{near-Carrollian diffeomorphism}\footnote{Expansion in $c^2$ has been dubbed pre-ultra-local expansion in \cite{Hansen:2021fxi}.}.

As stated previously, we need to find how the metric components vary under the near-Carrollian diffeomorphism. To carry out this task, we employ the technology of the anomaly operator $\Delta_\xi$ which compares the spacetime transformtaion of the field to its field space transformation. The metric $h$ is covariant under the near-horizon diffeomorphism, meaning that its anomaly $\Delta_\xi h := \delta_\xi h - \cL_\xi h$ vanishes. The anomaly of the metric $h$ decomposes as
\begin{equation}
\begin{aligned}
\Delta_\xi h &= -2c^2 (\Delta_\xi \bm{k}) \circ \bm{k} + \Delta_\xi q \\
&= -2 c^2 (\iota_\ell\Delta_\xi \bm{k}) \bm{k} \circ \bm{k} + 2\left( \Delta_\xi q (\ell,e_A) - c^2 (\iota_{e_A}\Delta_\xi \bm{k})\right) \bm{k} \circ \bm{e}^A  + \Delta_\xi q (e_A,e_B) \bm{e}^A \circ \bm{e}^B.
\end{aligned}
\end{equation}
Demanding covariance, $\Delta_\xi h =0$, imposes the following conditions,
\begin{align}
\iota_\ell\Delta_\xi \bm{k} = 0, \qquad  \Delta_\xi q (\ell,e_A) = c^2 (\iota_{e_A}\Delta_\xi \bm{k}) , \qquad \text{and} \qquad \Delta_\xi q (e_A,e_B) =0. 
\end{align}
The problem then boils down to the computation of the anomaly of the Ehresmann connection $\bm{k}$ and the anomaly of the null Carrollian metric $q$ (we defer the derivations to the Appendix \ref{fluid-transf}). Solving the above conditions for different powers of $c^2$ gives us the transformation of the metric components under the near-Carrollian diffeomorphism, 
\begin{subequations}
\lb{transf-fluid}
\begin{align}
\bbdelta_\xi \alpha &=  \delta^\Carr_{(\tau,Y)} \alpha \\
\e^\a \bbdelta_\xi \beta_A &= \e^\a \delta^\Carr_{(\tau,Y)} \beta_A  + \mrq_{AB}\ell[Z^B]  \\
\bbdelta_\xi \mrq_{AB} &=\delta^\Carr_{(\tau,Y)} \mrq_{AB}  \\
\bbdelta_\xi \lam_{AB} &=  \frac{1}{2}\delta^\Carr_{(\psi,Z)} \mrq_{AB} + \tau \ell[\lam_{AB}] + Y^C \mrD_C \lam_{AB} + 2 \lam_{C(A}\mrD_{B)} Y^C,
\end{align}
\end{subequations}
where we recalled the functional form of the Carrollian transformations\footnote{Although now there is no constraint on $Y^A$, unlike the Carrollian transformations where $\ell[Y^A] =0$.} \eqref{del-C-a}, \eqref{del-C-b}, and \eqref{del-C-q}, and the transformation of the velocity field is given by,
\begin{align}
\bbdelta_\xi V^A = - D_u Y^A. \lb{transf-V}
\end{align}

\subsection{Hydrodynamics from Near-Carrollian Diffeomorphism}

The Carrollian hydrodynamic equations \eqref{E-eq}, \eqref{C-eq}, \eqref{P-eq}, and \eqref{J-eq} can be recovered by demanding invariance, up to boundary terms, of the fluid action \eqref{S-fluid} under the near-Carrollian transformations, $\delta_\xi S_{\text{fluid}} =0$. Using the near-Carrollian transformations \eqref{transf-fluid} and \eqref{transf-V} and the Stokes theorem \eqref{Stokes}, one can show that
\begin{equation}
\begin{aligned}
\delta_\xi S_{\text{fluid}} = -\int_H \left( \tau \mathbb{E} +\bar{\psi} \mathbb{C} + Y^A \mathbb{P}_A + \bar{Z}^A\mathbb{J}_A \right) \mr{\bm{\epsilon}}_H + \Delta Q_\xi  \label{diff1}
\end{aligned}
\end{equation}
where we defined the combinations of the transformation parameters, $\bar{\psi} := \psi +\lam\tau$ and $\bar{Z}^A := Z^A + \lam Y^A$. The boundary term $\Delta Q_\xi$ is the difference of  Noether charges corresponding to the near-Carrollian diffeomorphism at the two ends of $H$. We clearly see that imposing $\delta_\xi S_{\text{fluid}} =0$ up to the boundary term yields the fluid equations. 

The Noether charges of these transformations have three components associated with different sectors of the near-Carrollian symmetries, 
\begin{equation}
\begin{aligned}
Q_\xi= Q_\tau + Q_Y + Q_{\bar{Z}},
\end{aligned}
\end{equation}
where each components are given by
\begin{subequations}
\begin{align}
Q_\tau &= -\int_S \tau \left( \sE + \e^\a \sJ^A \beta_A\right) \mr{\bm{\epsilon}}_S, \\
Q_Y  & = \int_S Y^A \left( \pi_A + \e^\a \left( \sK_A{}^B+\sP\delta_A^B \right)\beta_B\right) \mr{\bm{\epsilon}}_S, \\
Q_{\bar{Z}} &= \int_S \bar{Z}^A \left( \sJ_A + \e^\a \Sigma_A{}^B\beta_B\right) \mr{\bm{\epsilon}}_S.
\end{align}
\end{subequations}
where $S$ is a sphere at constant $u$.\footnote{One can more generally express the charges at the spheres $S_f= \{ u= f(\sigma)\}$ as the same integrals with $\beta_A $ replaced by $\beta_A -e_A[f]$.}
As one would expect, the transformations labelled by $\bar{\psi}$ has zero Noether charges, as they are generators of the non-dynamical constraint \eqref{C-eq}. This means that the $\bar{\psi}$ are pure gauge.

It is important to appreciate that our results generalize those presented in \cite{Ciambelli:2018ojf} (which was only the case $V^A =0$ and $\lam_{AB} =0$). In our consideration, we allow non-zero $V^A$ and $\lam_{AB}$ and by using the proposed near-Carrollian diffeomorphism \eqref{fluid-diff}, we managed to derive the complete set of Carrollian hydrodynamic equations and identified all the Noether charges. 

One can then compute the evolution of the charges. For the component $Q_\tau[u]$, we straightforwardly evaluate its time evolution using the energy equation \eqref{E-eq}, 
\begin{equation}
\begin{aligned}
\frac{\rd}{\rd u} Q_\tau & = - \int_S \bigg[ \e^\a \left( \tau (\ell + \mr{\theta})[\sE] + \sE \ell[\tau] \right) + \frac{1}{\sqrt{\mrq}} D_u ( \sqrt{\mrq}\tau \e^\a \sJ^A \beta_A) \bigg]\mr{\bm{\epsilon}}_S \\
&= \int_S (\tau \e^\a  \mathbb{E}) \mr{\bm{\epsilon}}_S + \int_S \e^\a \left( - \sE \ell[\tau] -\sJ^A (e_A - \varphi_A)[\tau] + \tau (\sS^{AB} + \sP \mrq^{AB})\mr{\theta}_{AB} + \tau\Sigma^{AB} \ell[\lam_{AB}] \right)\mr{\bm{\epsilon}}_S \\
&= \int_S (\tau \e^\a  \mathbb{E}) \mr{\bm{\epsilon}}_S + \int_S \e^\a \left( - \sE \bbdelta_\tau \alpha + \e^\a \sJ^A \bbdelta_\tau \beta_A + \frac{1}{2}(\sS^{AB} + \sP \mrq^{AB})\bbdelta_\tau \mrq_{AB} + \Sigma^{AB} \bbdelta_\tau \lam_{AB} \right)\mr{\bm{\epsilon}}_S
\end{aligned}
\end{equation}
More generally one finds that the charge evolution equations  can be written as
\begin{equation}
\begin{aligned}
\frac{\rd}{\rd u} Q_\xi
=\ & \int_S \e^\a \left( \tau \mathbb{E} +\bar{\psi} \mathbb{C} + Y^A \mathbb{P}_A + \bar{Z}^A\mathbb{J}_A \right) \mr{\bm{\epsilon}}_S \\
& + \int_S \e^\a \left( - \sE \bbdelta_\xi \alpha + \e^\a \sJ^A \bbdelta_\xi \beta_A - \e^{-\a} \tilde{\pi}_A \bbdelta _\xi V^A + \frac{1}{2}(\tilde{\sS}^{AB} + \sP \mrq^{AB})\bbdelta_\xi \mrq_{AB} + \Sigma^{AB} \bbdelta_\xi \lam_{AB} \right)\mr{\bm{\epsilon}}_S.
\end{aligned}
\end{equation}
These equations can be derived directly from combining \eqref{diff1} with \eqref{S-fluid}.

\section{Conclusion} \lb{conclusion}

In this work we have extended the analysis of the $c\to0$ limit of relativistic fluids towards a Carrollian fluids. Starting from Carroll structures, we studied Carrollian fluids and presented two methods to derive the corresponding hydrodynamic conservation laws. In the first and conventional method, we started from energy-momentum tensors of general relativistic fluids, then properly consider the Carrollian limit ($c \to 0$) of the standard relativistic conservation laws. Our derivations could be viewed as the generalization of \cite{Ciambelli:2018xat} due to the fact that we now have in our construction the fluid velocity $V^A$ and the sub-leading components of sphere metric $\lam_{AB}$. These two quantities are indeed important parts of the phase space of Carrollian hydrodynamics. The second route, which was the highlight of this article, was to viewed Carrollian hydrodynamics as the consequence of symmetries. We argued that Carrollian diffeomorphism is not sufficient to derive the full set of Carrollian fluid equations (which has already been studied in \cite{Ciambelli:2018ojf, Petkou:2022bmz}) and that we need to go beyond Carrollian diffeomorphism. To this end, we introduced the notion of near-Carrollian symmetries \eqref{fluid-diff} and finally showed that it leads to the complete set of Carrollian hydrodynamic equations.

Many directions however remain to be explored. Let us list some of them below.

\begin{enumerate}[label=\roman*)]

\item \emph{Realization on stretched horizons and null boundaries}: The membrane paradigm \cite{Damour:1978cg, Thorne:1986iy, Price:1986yy} has established the correspondence between black hole physics and dynamics of fluids living on timelike surfaces, called stretched horizons or membranes, located near black hole horizons (which are null surfaces). As Carroll structures are universal structures of null boundaries, be they at finite distances \cite{Chandrasekaran:2018aop, Chandrasekaran:2021hxc, Ashtekar:2021wld} or infinities \cite{Ashtekar:2014zsa, Ashtekar:2018lor}, one would therefore expect the membrane fluids to be Carrollian fluids. This statement has just been realized recently in \cite{Donnay:2019jiz} (see also \cite{Penna:2018gfx}), where it has been shown that the Einstein equations on black hole horizons can be displayed as Carrollian hydrodynamic equations and that the near-horizon diffeomorphism \cite{Donnay:2015abr, Donnay:2016ejv} is Carrollian diffeomorphism. The analog of the Brown-York energy-momentum tensor of null boundaries and its conservation laws have also been studied in \cite{Chandrasekaran:2021hxc}. 

We have learned in this work that Carroll structures can be endowed on any surfaces, regardless of whether they are null or timelike\footnote{Usually in the literature, the Carrollian metric $q$ is treated as an induced metric on hypersurfaces and the null-ness property of $q$ then dictates the hypersurfaces to be null. This is not necessary as we can endowed, for example, the Lorentzian metric \eqref{RPmetric} on any type of hypersurfaces while incorporating all elements of the Carroll structure into the geometry of the surfaces.}, inferring the possibility to assign the Carrollian hydrodynamic picture to timelike surfaces, say for example, stretched horizons. In fact, it is to be expected that stretched horizons encode some underlying informations of the null boundaries, in the same spirit as the near-Carrollian analysis (the value of $c^2$ deviates from zero) presented in this work. To make our argument more elaborate, further investigations are required and some aspects of it will be provided in our upcoming work \cite{Jaiakson:2022}.

\item \emph{Sub-subleading and higher order corrections}: In our analysis, only the sub-leading (order $c^2$) terms were considered. One can indeed extend our construction by including sub-subleading (order $c^4$) and higher order terms in the metric \eqref{RPmetric}, which will introduce new variables to the phase space of Carrollian fluids and in turn activates new Carrollian fluid momenta conjugate to these higher-order variables. These momenta are corresponding to the $c^{-4}, c^{-6}, c^{-8}, ... $ corrections of the dissipation tensors \eqref{q-tau} which we have truncated them at order $c^{-2}$ here. As a consequence, the near-Carrollian diffeomorphism \eqref{fluid-diff} will be enhanced with the inclusion of higher-order corrections associated with new equations governing the dynamics of these new momenta and also new Noether charges. 

Let us also mention that this picture has already been realized in the context of asymptotic null infinities \cite{Freidel:2021qpz, Freidel:2021dfs, Freidel:2021ytz} which exhibit the infinite tower of charges and their corresponding conservation equations. It would then be of interest to study the higher-order dynamics of the Carrollian hydrodynamics and bridge the findings with the results at infinities.

\item \emph{Thermodynamics of Carrollian fluids}: Having established the Carrollian hydrodynamics, one natural question therefore emerges --- \emph{what are thermodynamical properties of Carrollian fluids?} Admittedly, although this question may not garner much interest in the field of fluid mechanics due to the sole fact that everyday life's fluids are Galilean in nature, we believe that answering this question will provide useful insights to the realm of black hole physics. One possible direction to explore in the future is the notion of \emph{thermodynamical horizons}, the type of surfaces that obey all laws of thermodynamics, and also the universal notion of equilibrium in any surface. 

\item \emph{Galilean hydrodynamics from symmetries}: As the speed of light $c$ now plays a role of the varying parameter when taking non-Lorentzian limits, it then suggests that similar analysis could be carried out for the Galilean case ($c \to \infty$ limit), therefore giving the derivation of Galilean hydrodynamics, e.g., the continuity equation and the Navier-Stokes equations, from symmetries. In this case, the underlying structure is the Newton-Cartan structure \cite{Duval:2014uoa} (see also \cite{Duval:1984cj, Duval:1990hj}) instead of the Carroll structure.

\end{enumerate}

\section*{Acknowledgments}
We would like to thank C\'eline Zwickel for helpful discussions and insights. Research at Perimeter Institute is supported in part by the Government of Canada through the Department of Innovation, Science and Economic Development Canada and by the Province of Ontario through the Ministry of Colleges and Universities. The work of LF is funded by the Natural Sciences and Engineering Research Council of Canada (NSERC) and also in part by the Alfred P. Sloan Foundation, grant FG-2020-13768. This project has received funding from the European Union's Horizon 2020 research and innovation programme under the Marie Sklodowska-Curie grant agreement No. 841923. PJ's research is supported by the DPST Grant from the government of Thailand, and Perimeter Institute for Theoretical Physics.


\appendix

\section{Coordinate expressions for $\varphi_A$ and $w_{AB}$} \lb{Phi-W-derive}

Expressions for the Carrollian acceleration $\varphi_A$ and the Carrollian vorticity in coordinates are straightforwardly computed from the Carrollian commutators. Let us start with the acceleration, we evaluate 
\begin{equation}
\begin{aligned}
\varphi_A \ell &= [\ell, e_A] \\
&= [\e^{-\a} D_u, e_A]\\
&= e_A[\alpha]\ell + \e^{-\alpha}[ D_u, (J^{-1})_A{}^B\pa_B +\beta_A D_u] \\
&= \left( D_u \beta_A +e_A[\alpha]  \right)\ell + \e^{-\alpha} \left( D_u(J^{-1})_A{}^B - (J^{-1})_A{}^C\pa_C V^B  \right)\pa_B.
\end{aligned}
\end{equation}
The last term vanishes due to the condition \eqref{Carrollian}. We therefore obtain the expression
\begin{align}
\varphi_A =  D_u \beta_A +e_A[\alpha]. 
\end{align}
Similarly, the Carrollian vorticity can be evaluated as follows,
\begin{equation}
\begin{aligned}
w_{AB} \ell &= [e_A, e_B] \\
&= [(J^{-1})_A{}^C\pa_C +\beta_A D_u, (J^{-1})_B{}^D\pa_D +\beta_B D_u] \\
&= [(J^{-1})_A{}^C\pa_C , (J^{-1})_B{}^D\pa_D ] + [(J^{-1})_A{}^C\pa_C , \beta_B D_u] +[\beta_A D_u, (J^{-1})_B{}^D\pa_D] \\
& \ \ \ \ + [\beta_A D_u, \beta_B D_u] \\
& = \e^\a \left( e_A[\beta_B] - e_B[\beta_A] \right) \ell + \left( e_A[J_B{}^C] - \beta_A (J^{-1})_B{}^D\pa_D V^C - (A \leftrightarrow B)\right)\pa_C
\end{aligned}
\end{equation}
The last term, again, computes to zero by means of \eqref{Carrollian}. The Carrollian vorticity is then given by
\begin{align}
w_{AB} = \e^\a \left( e_A[\beta_B] - e_B[\beta_A] \right).
\end{align}
One can alternatively check by computing the curvature of $\bm{k} = \e^\a (\bm{\rd} u - \beta_A \bm{e}^A)$, which is
\begin{equation}
\begin{aligned}
\bm{\rd k} &= \bm{\rd} \alpha \wedge \bm{k} - \e^\a \bm{\rd} \beta_A \wedge \bm{e}^A \\
&= - \left(  D_u \beta_A +e_A[\alpha]\right) \bm{k} \wedge \bm{e}^A - \frac{1}{2} \e^\a \left( e_A[\beta_B] - e_B[\beta_A] \right) \bm{e}^A \wedge \bm{e}^B \\
&= - \varphi_A \bm{k} \wedge \bm{e}^A - \frac{1}{2} w_{AB} \bm{e}^A \wedge \bm{e}^B. 
\end{aligned}
\end{equation}


\section{Horizontal covariant derivative} \lb{hor-derivative}


One property of the horizontal covariant derivative $\sD_A$ is that we can define the analog of the Riemann tensor with this connection and it is called the Riemann-Carroll tensor, ${}^{\sss (2)}R^A{}_{BCD}$. Its components are determined from the commutator,
\begin{align}
[\sD_C, \sD_D] X^A = \two R^A{}_{BCD} X^B + w_{CD} \ell[X^A], \lb{Rie-Car}
\end{align}
where the vertical derivative term $\ell[X^A]$ appeared due to the non-integrability of the horizontal subspace.   
We can then define corresponding the Ricci-Carroll tensor, ${}^{\sss (2)}R_{AB} :=  {}^{\sss (2)}R_{CADB} q^{CD}$, and the Ricci-Carroll scalar, ${}^{\sss (2)}R :=  {}^{\sss (2)}R_{AB} q^{AB}$. Let us also note that the Ricci-Carroll tensor is not symmetric, $\two R_{AB} \neq \two R_{BA}$, in general. 

Since we are dealing with the expansion in $c^2$ of the sphere metric, $q_{AB} = \mrq_{AB} + 2c^2 \lam_{AB}$, it then becomes essential to define the similar expansion for the connection $\two \Gamma^A_{BC}$. With this in mind, let us define the following connection, 
\begin{align}
{}^{\sss (2)}\mr{\Gamma}^A_{BC} := \frac{1}{2} \mrq^{AD}\left( e_B[ \mrq_{DC}] +e_C[ \mrq_{BD}] - e_D [\mrq_{BC}] \right),
\end{align}
and the new horizontal covariant derivative $\mrD_A$ compatible with the zeroth-order of the sphere metric $\mrq_{AB}$, that is $\mrD_A \mrq_{BC} =0$. This operator $\mrD_A$ acts on a horizontal tensor the same way as $\sD_A$ but with the new connection $\two \mr{\Gamma}^A_{BC}$ instead of $\two \Gamma^A_{BC}$. One can therefore show that $\two \Gamma^A_{BC}$ admits the following expansion in $c^2$,
\begin{align}
\two \Gamma^A_{BC} =  \two\mr{\Gamma}^C_{AB} + c^2 \left( \mrD_A\lam_B{}^C +\mrD_B\lam_A{}^C-\mrD^C\lam_{AB} \right) + \O(c^4).
\end{align}


\section{Integration by parts}\label{intbypart}
One can verify the following relations
\begin{align}
\bm{\rd}(f \bm{\epsilon}_S) = \left( \ell[f] + \theta f\right) \bm{\epsilon}_H, \qquad \text{and}  \qquad \bm{\rd} \left( \iota_{X} \bm{\epsilon}_H \right) = \left( \sD_A X^A +\varphi_A X^A \right) \bm{\epsilon}_H,
\end{align}
for a function $f$ on $H$ and for a horizontal vector $X=X^A e_A \in \text{\bf hor}(H)$. The second equation can be proven as follows:
\begin{equation}
\begin{aligned}
\bm{\rd} (\iota_X \bm{\epsilon}_H) = \cL_X \bm{\epsilon}_H &= \left( X[\ln \sqrt{q}] + \left( \iota_\ell \cL_X \bm{k} \right) +  \left( \iota_{e_A} \cL_X \bm{e}^A \right)  \right)\bm{\epsilon}_H \\
&= \left( e_A[X^A] + X^A e_A[\ln \sqrt{q}] + \varphi_A X^A  \right)\bm{\epsilon}_H \\
&= \left( \sD_A X^A +\varphi_A X^A \right) \bm{\epsilon}_H,
\end{aligned}
\end{equation}
where we recalled the expression of the volume form $\bm{\epsilon}_H = \frac{1}{2} \varepsilon_{AB} \sqrt{q} \ \bm{k} \wedge \bm{e}^A \wedge \bm{e}^B$ and the curvature of the Ehresmann connection \eqref{d-k}. 

One can imagine the space $H$ to have a boundary $\partial H$ situated at a constant value of the coordinate $u$. This boundary, in our construction, is identified under the projection map with the sphere $S$, meaning that $\pa H = S_u$. In this setup, the Stokes theorem is written as
\begin{subequations}
\label{Stokes}
\begin{align}
\int_H \left( \ell[f] + \mr{\theta} f\right) \mr{\bm{\epsilon}}_H &= \int_{S_u} f \mr{\bm{\epsilon}}_S, \\
\int_H \left( \mrD_A X^A +\varphi_A X^A \right) \mr{\bm{\epsilon}}_H &= \int_{S_u} \e^\a X^A \beta_A \mr{\bm{\epsilon}}_S.
\end{align}
\end{subequations}

Alternatively, one can choose a more general cut, say $u = f(\s)$. In this case, we have instead 
\begin{align}
\int_H \left( \mrD_A X^A +\varphi_A X^A \right) \mr{\bm{\epsilon}}_H &= \int_{S_f} \e^\a X^A (\beta_A - e_A[f]) \mr{\bm{\epsilon}}_S.
\end{align}
Let us observe that choosing $\beta_A = e_A[f]$ removes the boundary contribution. This is equivalent to the case of vanishing vorticity, $w_{AB} =0$.

\section{Covariant derivatives} \lb{cov-deri}

In the main text, we already presented the derivation of the covariant derivative $\nabla_\ell \ell$. Here we complete the detailed derivations of the remaining covariant derivatives, which are $\nabla_{e_A} \ell, \nabla_{\ell} e_A$, and $\nabla_{e_A} e_B$. 

\begin{itemize}[leftmargin=*]

\item \textbf{Derivation of $\nabla_\ell \ell$:} For completeness, let us quote the result derived in the main text,
\begin{align}
\nabla_\ell \ell  = c^2 \varphi^A e_A + \mathcal{\O}(c^4).
\end{align}
\item \textbf{{Derivation of $\nabla_{e_A} \ell$}:} We begin by writing the vector $\nabla_{e_A} \ell$ in the Carrollian basis $(\ell, e_A)$, 
\begin{align}
\nabla_{e_A} \ell = \left(k_i\nabla_{e_A} \ell^i \right) \ell +\left(e^B{}_i\nabla_{e_A} \ell^i \right)e_B,
\end{align}
then consider each component separately. The vertical component is identically zero as one can easily see from
\begin{align}
k_i\nabla_{e_A} \ell^i  = -\frac{1}{c^2}h(\ell, \nabla_{e_A} \ell) = -\frac{1}{2c^2}e_A \left[h(\ell, \ell) \right] =0.
\end{align}
The horizontal components are computed using repeatedly the Leibniz rule and the commutators \eqref{C-comm}, 
\begin{equation}
\begin{aligned}
e^B{}_i\nabla_{e_A} \ell^i &= q^{BC}h(e_C, \nabla_{e_A} \ell) \\
&= \frac{1}{2}q^{BC} \left( h(e_C, \nabla_{e_A} \ell) + h(e_C, \nabla_{e_A} \ell) \right) \\
&= \frac{1}{2}q^{BC} \left( -h(\nabla_{e_A}e_C,  \ell) + h(e_C, \nabla_{e_A} \ell) \right) \\
&= \frac{1}{2}q^{BC} \left( -h([e_A,e_C],\ell)-h(\nabla_{e_C}e_A,  \ell) + h(e_C, \nabla_{e_A} \ell) \right) \\
&= \frac{1}{2}q^{BC} \left( c^2w_{AC}+h(e_A, \nabla_{e_C} \ell) + h(e_C, \nabla_{e_A} \ell) \right) \\
&= \frac{1}{2}q^{BC} \left( c^2w_{AC}+h(e_A, \nabla_\ell e_C) + h(e_C, \nabla_\ell e_A) \right) \\
&= \frac{1}{2}q^{BC} \left( c^2w_{AC}+2\theta_{AC} \right), \\
\end{aligned}
\end{equation}
where we recalled $2\theta_{AB}  = \ell[q_{AB}]$. Expanding the metric $q_{AB}$ in $c^2$, we therefore obtain 
\begin{align}
\nabla_{e_A} \ell= \left(\mr{\theta}_A{}^B + c^2 \left(\frac{1}{2}w_A{}^B + \mrq^{BC} \ell[\lam_{AC}] - 2 \lam^{BC} \mr{\theta}_{AC} \right)\right)e_B + \O(c^4).
\end{align}
\item \textbf{Derivation of $\nabla_\ell e_A$:} This term decomposes as
\begin{align}
\nabla_\ell e_A = \left( k_i \nabla_\ell e_A{}^i\right) \ell + \left( e^B{}_i \nabla_\ell e_A{}^i\right) e_B.
\end{align}
Its components are already determined by the components of $\nabla_\ell \ell$ and $\nabla_{e_A} \ell$. For the vertical component, we have
\begin{align}
k_i \nabla_\ell e_A{}^i = -\frac{1}{c^2} h(\ell, \nabla_\ell e_A) = \frac{1}{c^2} h(\nabla_\ell \ell, e_A) = \varphi_A,
\end{align}
and for the horizontal components, we have
\begin{equation}
\begin{aligned}
e^B{}_i \nabla_\ell e_A{}^i &= q^{BC} h(e_C,\nabla_\ell e_A) \\
&=q^{BC} \left(h(e_C,\nabla_{e_A} \ell)+ h(e_C, [\ell,e_A])  \right) \\
& = \mr{\theta}_A{}^B + c^2 \left( \frac{1}{2}w_A{}^B + \mrq^{BC} \ell[\lam_{AC}] - 2 \lam^{BC} \mr{\theta}_{AC} \right) + \O(c^4).
\end{aligned}
\end{equation}
Together, they give
\begin{align}
\nabla_\ell e_A = \varphi_A \ell+\left(\mr{\theta}_A{}^B + c^2 \left( \frac{1}{2}w_A{}^B + \mrq^{BC} \ell[\lam_{AC}] - 2 \lam^{BC} \mr{\theta}_{AC} \right)\right)e_B + \O(c^4).
\end{align}

\item \textbf{Derivation of $\nabla_{e_A} e_B$:} For this covariant derivative, we write its decomposition in the Carrollian basis as
\begin{align}
\nabla_{e_A} e_B = \left( k_i\nabla_{e_A} e_B{}^i \right) \ell + \left( e^C{}_i\nabla_{e_A} e_B{}^i \right)e_C,
\end{align}
where the vertical component is 
\begin{align}
k_i\nabla_{e_A} e_B{}^i = -\frac{1}{c^2}h(\ell, \nabla_{e_A} e_B) = \frac{1}{c^2}h(\nabla_{e_A}\ell,  e_B) = \frac{1}{c^2} \mr{\theta}_{AB} + \left(\frac{1}{2}w_{AB} + \ell[\lam_{AB}]\right). 
\end{align}
The horizontal components, $e^C{}_i\nabla_{e_A} e_B{}^i = q^{CD}h(e_D, \nabla_{e_A} e_B)$, can be evaluated using the following trick. First, we use that the covariant derivative is metric compatible, which following from this the obvious identity, $e_A[q_{DB}] = h(e_D, \nabla_{e_A} e_B) +h(e_B, \nabla_{e_A} e_D)$. It then become a straightforward computation to show that
\begin{equation}
\begin{aligned}
e_A[q_{DB}] +e_B[q_{AD}]-e_D[q_{AB}] =& \ 2h(e_D, \nabla_{e_A}e_B) + h(e_A, [e_B, e_D]) \\ 
&+ h(e_B, [e_A, e_D]) +h(e_D, [e_B, e_A]).
\end{aligned}
\end{equation}
Using the commutator $[e_A, e_B] = w_{AB} \ell$ and that $h(e_A,\ell) =0$, we arrive at the expression for the horizontal components,
\begin{equation}
\begin{aligned}
e^C{}_i\nabla_{e_A} e_B{}^i &= \frac{1}{2} q^{CD}\left( e_A[q_{DB}] +e_B[q_{AD}]-e_D[q_{AB}]\right) = \two\Gamma^C_{AB}.
\end{aligned}
\end{equation}
We finally obtain the covariant derivative $\nabla_{e_A} e_B$ expanded in $c^2$ as
\begin{equation}
\begin{aligned}
\nabla_{e_A} e_B = \ & \left(\frac{1}{c^2} \mr{\theta}_{AB} + \left(\frac{1}{2}w_{AB} + \ell[\lam_{AB}]\right) \right) \ell + \two\mr{\Gamma}^C_{AB} e_C \\
&+c^2 \left( \sD_A\lam_B{}^C +\sD_B\lam_A{}^C-D^C\lam_{AB} \right)e_C.
\end{aligned}
\end{equation}

\end{itemize}


\section{Anomaly computations}  \lb{fluid-transf}

To evaluate the anomaly of the Ehresmann connection, $\Delta_\xi \bm{k} = \delta_\xi \bm{k} - \cL_\xi \bm{k}$, one first computes its variation under the near-Carrollian diffeomorphism. Using the fact that the the coordinates $x^i = (u,y^A)$ are field-independent and thus $\delta \bm{\rd}x^i = 0$, we can straightforwardly write the variation of the Ehresmann connection as
\begin{equation}
\begin{aligned}
\delta_\xi \bm{k} 
&= \bbdelta_\xi \alpha  \bm{k} - \e^\a \bbdelta_\xi \beta_A  \bm{e}^A.
\end{aligned}
\end{equation} 
Next, we need to compute the Lie derivative of the Ehresmann connection. Using the Cartan formula and recalling the curvature of the Ehresmann connection \eqref{d-k}, one can proof that 
\begin{equation}
\begin{aligned}
\cL_\xi \bm{k} &=\bm{\rd}(\iota_\xi \bm{k}) +\iota_\xi \bm{\rd k} \\
&= \bm{\rd} f + (X \cdot \varphi) \bm{k} + \left( - f \varphi_A + w_{AB} X^B \right) \bm{e}^A \\
&= (\ell[f] + X \cdot \varphi) \bm{k} + \left( (e_A-  \varphi_A)[f] +w_{AB} X^B  \right). \bm{e}^A
\end{aligned}
\end{equation}
Expanding the transformation parameter $f = \tau + c^2 \psi$ and $X^A = Y^A + c^2 Z^A$, the anomaly of the Ehresmann connection $\Delta_\xi \bm{k}$ decomposes as
\begin{align}
\Delta_\xi  \bm{k} = \left(\iota_\ell \Delta_\xi \bm{k} \right) \bm{k} + \left(\iota_{e_A} \Delta_\xi \bm{k} \right) \bm{e}^A,
\end{align}
where the components are 
\begin{equation}
\begin{aligned}
\iota_\ell \Delta_\xi \bm{k} 
& = \bbdelta_\xi \alpha - \bbdelta^\Carr_{(\tau,Y)} \alpha  + \O(c^2),\\
\iota_{e_A} \Delta_\xi \bm{k} &= -\e^\a \left( \bbdelta_\xi \beta_A -\bbdelta^\Carr_{(\tau,Y)} \beta_A  \right) +\O(c^2). 
\end{aligned}
\end{equation}
Next, we compute the anomaly of the null Carrollian metric, $q= q_{AB} \bm{e}^A \circ \bm{e}^B$. We begin by considering its variation under the near-Carrollian diffeomorphism and show that
\begin{equation}
\begin{aligned}
\delta_\xi q & = -2\e^{-\a}\left(q_{AB} \bbdelta_\xi V^B \right) \bm{k} \circ \bm{e}^A + \bbdelta_\xi q_{AB} \bm{e}^A \circ \bm{e}^B.
\end{aligned}
\end{equation}
Using the Cartan formula and the fact that $\bm{\rd e}^A =0$, the Lie derivative of the null Carrollian metric thus given by
\begin{equation}
\begin{aligned}
\cL_\xi q &= \xi[q_{AB}] \bm{e}^A \circ \bm{e}^B + 2 q_{AB} (\cL_\xi \bm{e^A}) \circ \bm{e}^B \\
&= 2 q_{AB} \ell[X^B] \bm{k}\circ \bm{e}^B + \left( \xi[q_{AB}] + q_{C (A} e_{B)}[X^C] \right) \bm{e}^A \circ \bm{e}^B.
\end{aligned}
\end{equation}
The anomaly of the null Carrollian metric is 
\begin{align}
\Delta_\xi q = \delta_\xi q - \cL_\xi q = 2\Delta_\xi q(\ell, e_A) \bm{k} \circ \bm{e}^A + \Delta_\xi q(e_A, e_B) \bm{e}^A \circ \bm{e}^B,
\end{align}
where its components, given in the $c^2$-expansion, are given by 
\begin{equation}
\begin{aligned}
\Delta_\xi q(\ell, e_A) = -  \e^{-\a} (\mrq_{AB} + 2c^2 \lam_{AB}) \left(\bbdelta_\xi V^B + D_uY^B \right) - c^2 \mrq_{AB}  \ell[Z^B],
\end{aligned}
\end{equation}
and
\begin{equation}
\begin{aligned}
\Delta_\xi q(e_A,e_B) = \ &  \left(\bbdelta \mrq_{AB} - \bbdelta^\Carr_{(\tau,Y)} \mrq_{AB} \right)  \\
&+ 2c^2 \bigg( \bbdelta_\xi \lam_{AB} -  \frac{1}{2}\bbdelta^\Carr_{(\psi,Z)} \mrq_{AB}  - \tau \ell[\lam_{AB}] - Y^C \mrD_C \lam_{AB} - 2 \lam_{C(A}\mrD_{B)} Y^C \bigg). 
\end{aligned}
\end{equation}

\section{Carrollian fluid action} \lb{App:action}

The detail of the derivation of the Carrollian fluid action \eqref{S-fluid} is provided here. Recalling that the variation of the action is given by
\begin{align}
\delta S_{\text{fluid}} = \int_H \frac{1}{2}\left( T^{\sss (0)}{}^{ij} \delta h_{\sss(0)}{}_{ij} +T^{\sss (-1)}{}^{ij} \delta h_{\sss(1)}{}_{ij} + \lam T^{\sss (-1)}{}^{ij} \delta h_{\sss(0)}{}_{ij}\right)\mr{\bm{\epsilon}}_H, \lb{action-derive}
\end{align}
the derivation thus boils down the components of the relativistic energy-momentum tensor and the metric on $H$ in the $c^2$-expansion. 

For the energy-momentum tensor \eqref{rela-em}, we have that 
\begin{subequations}
\begin{align}
T^{\sss(-1)}{}^{ij} &= \sE \ell^i \ell^j + \sJ^A (e_A{}^i \ell^j +e_A{}^j \ell^i) + \Sigma^{AB} e_A{}^i e_B{}^j   \\
T^{\sss(0)}{}^{ij} &= (\sS^{AB} + \sP \mr{q}^{AB})e_A{}^i e_B{}^j + (\pi^A - 2\lam^A{}_B \sJ^B) (e_A{}^i \ell^j + e_A{}^j \ell^i).
\end{align}
\end{subequations}
Using the results from Appendix \ref{fluid-transf}, we can show that the components of the expansion of the metric variation are
\begin{subequations}
\begin{align}
\delta h_{\sss(0)}{}_{ij} &= -\e^{-\a}\left(\mr{q}_{AB} \bbdelta V^B \right) (k_i e^A{}_j +k_j e^A{}_i  ) + (\bbdelta \mrq_{AB}) e^A{}_i e^B{}_j  \\
\delta h_{\sss(1)}{}_{ij} &= -(\bbdelta \alpha) k_i k_j + (\e^\a \bbdelta \beta_A - 2\e^{-\a} \lam_{AB} \bbdelta V^B )   (k_i e^A{}_j + k_j e^A{}_i)  + 2(\bbdelta \lam_{AB}) e^A{}_i e^B{}_j. 
\end{align}
\end{subequations}
Straightforward computations of \eqref{action-derive} yields the variation of the Carrollian fluid action \eqref{S-fluid} presented in the main text. 

\bibliography{Biblio.bib}
\bibliographystyle{Biblio}

\end{document}